\documentclass[12pt]{article}

\usepackage{amsfonts,amssymb}

\def\i{\mbox{i}}
\def\d{\mbox{d}}
\def\Im{\mbox{Im}}
\def\Re{\mbox{Re}}

\title{Modified perturbation theory for pair production
and decay of fundamental unstable particles}
\author{M. L. Nekrasov }

\date{}
\begin{document}
\maketitle

\begin{abstract}

We construct an asymptotic expansion in powers of the coupling
constant directly of the cross-section for pair production and
decay of fundamental unstable particles. The resonant and
kinematic singularities arising in the expansion we treat in the
sense of distributions. This mode allows us to transform formally
divergent integrals into absolutely convergent ones with keeping
the asymptotic property of the expansion. The appropriate
procedure is elaborated up to an arbitrary order of the expansion.
The peculiarity of application of the procedure in the threshold
region is analysed. The scheme of the calculations within the NNLO
approximation is explicitly discussed.

%\keywords{unstable particles; modified perturbation theory.}
\end{abstract}

%\ccode{11.15.Bt, 13.66.Jn} % 13.38.-b, 14.65.Ha

\section{Introduction}\label{int}

The processes of pair production and decay of fundamental unstable
particles such as the top quarks or $W$ bosons provide a powerful
tool for the determination of their masses and parameters of their
interaction vertices. At the LHC and at an International linear
collider (ILC) \cite{LHC-ILC,ILC} the mentioned determination is
planned to be carried out through measurings of the absolute value
of the cross-section and its dependence on the center-of-mass
energy, and of the angular distributions and spin correlations. At
the ILC, where the higher precision is projected, the
corresponding theoretical calculations must be carried out in many
cases with the two-loop accuracy in matrix elements
\cite{EW,Top-QCD}. The calculation of the cross-sections
thereafter must be carried out with the equivalent accuracy.
However, the calculation of the cross-sections is obstructed by
resonant contributions of the unstable particles since such
contributions are nonintegrable in the framework of conventional
perturbation theory~(PT).

The common practice of a solution to this problem consists in the
Dyson resummation of the imaginary parts of self-energies in the
denominators of unstable-particles propagators. Simultaneously
this cures the Coulomb singularities, the other problematic
contributions in the cross-sections \cite{Fadin1}--\cite{Fadin4}.
However, the application of the Dyson resummation is rather
dangerous in gauge theories because of risk to break the gauge
cancellations necessary for unitarity. For this reason the
one-loop calculations at LEP2 were carried out in the double-pole
approximation (DPA), which guaranteed the gauge cancellations
\cite{LEP2,4f}. However in the ILC case the accuracy of DPA is no
longer sufficient \cite{CMS1,CMS2}. The application of the higher
orders of the ``pole expansion'' is ineffective, too, because of
systematic inaccuracy inherent in this approach (see discussion in
\cite{4f} and in the references therein). So in the ILC case the
methods are required that can provide completely systematic
calculations.

Among such methods as most perspective ones long time the methods
were considered that were based on the background-field and
pinch-technique formalisms. Both they imply a reconstruction of
the conventional PT series; through the formation of the
gauge-invariant effective action or through the Dyson resummation
after fulfilling the gauge cancellations (see
Refs.~\cite{BFM1,BFM2} and \cite{pinch1}--\cite{pinch4},
respectively, and the references therein). Unfortunately some
difficulties are inherent in the mentioned methods, too. In
particular, in the background-field formalism a dependence on the
quantum gauge parameter remains, which cannot be fixed on physical
grounds. The pinch-technique method does not have this problem,
but there is another difficulty which is the common one in both
cases. Namely for reaching the $O(\alpha^n)$ precision without
violating the gauge cancellation in the framework of these methods
it is not enough to calculate corrections up to the $O(\alpha^n)$
order; it is necessary to calculate {\it all} the
$O(\alpha^{n+1})$ loop corrections, too, at least their
imaginary-part contributions, which is impractical (see, in
particular, discussion in \cite{Ditt}).

Another prominent approach is the ``complex-mass scheme'' (CMS),
first considered in \cite{CMS01,CMS02} and then worked out in
\cite{CMS1,CMS2}. Actually, CMS is a renormalization scheme
defined through the identification of the renormalized mass of an
unstable particle with the complex pole of its full propagator. So
in the CMS the renormalized mass absorbs the imaginary
contributions of the conventional self-energy, becoming thus a
complex quantity. Unfortunately, this triggers the complex-valued
counterterms, which violates unitarity. However, at the one-loop
level the unitarity-violating contributions manifest themselves
effectively at the higher orders with the gauge cancellations
being explicitly maintained \cite{CMS1,CMS2}. So the one-loop
calculations get legitimate. Whether the same behavior takes place
at the two-loop level, this is not clear. For this reason the CMS
cannot be considered as a rigorous procedure \cite{CMS2}, at least
at present. So a development of alternative approaches still is an
actual problem.

In this paper we consider a modified perturbation theory (MPT),
first proposed in \cite{F1} and further discussed in
\cite{F2}--\cite{N-tt}. The basic idea of this approach is a
systematic expansion in powers of the coupling directly of the
probability instead of the amplitude. This mode allows one to
impart the sense of distributions \cite{Distr} to the propagators
squared of the unstable particles, and on this basis to
asymptotically expand the propagators squared without the
appearance of the divergences in the cross-section.
(Simultaneously, the rest of the amplitude squared is to be
expanded in the conventional sense.) In doing so one gets the MPT
expansion of the cross-section.

It must be emphasized that the appearance of the divergences in
the cross-section when it is considered as an integral of the
expanded amplitude squared, in fact does not yet mean
impossibility to expand the cross-section. Immediately this means
only an invalidity of the expansion of the integrand (the
amplitude squared), because in many cases the expansion becomes
possible after the calculation of the integral. In such cases the
distributions method is found to be a power tool to do that in
advance, before explicit calculation of the integral. The
advantage of the distributions method is caused by its possibility
to make divergent contributions convergent by basing on the
extension principle. Unfortunately, this procedure is accompanied
by ambiguities emerging in the integration rules. However, the
ambiguities may be reduced to the sum composed of the
$\delta$-functions and of their derivatives with arbitrary
coefficients, which should be added to the integrand
\cite{Distr,Gelfand}. So the ambiguities can be eliminated via the
fixing of the coefficients only. (Let us remember that the similar
phenomenon occurs in the case of the UV renormalization at the
rigid fixing of the renormalization scheme \cite{B-Sh}). In the
context of asymptotic expansion the coefficients may be fixed by
means of establishing the asymptotic property of the expansion
that should appear after the integration is carried out
\cite{Gelfand,F3,F4}. Furthermore, in some cases the expansion may
take the form of a complete series in powers of the expansion
parameter. At the expansion of the cross-section this would mean
that each term of the expansion is proportional to a power of the
coupling constant $\alpha$ and contains no other dependence on
$\alpha$. The latter property is extremely important since it
means that in the cross-section the gauge cancellations must take
place automatically at each power of $\alpha$ because the exact
solution is gauge-invariant.

At present it is generally known how to implement the MPT
expansion in the case of pair production and decay of unstable
particles, but up to the next-to-leading order (NLO) only and in
the context of models without massless particles \cite{N-tt}.
(This modelling, however, in practice has shown the existence of a
smooth line-shape for the observable cross-section in the
framework of MPT approach.) A further development of the method
should imply a construction of the cross-section in the higher
orders of the MPT expansion in realistic cases. Solving this
problem, one should overcome difficulties caused by the Coulomb
and kinematic singularities. Recall that the Coulomb singularities
arise as universal corrections caused by the exchanges by soft
massless particles (photons, gluons) between outgoing massive
particles in the limit of small relative velocities. Such
contributions are nonintegrable in the cross-section in the higher
orders of the conventional PT. However, there is a general method
of summing up the Coulomb singularities, which makes the result
integrable \cite{Fadin2}--\cite{Fadin4}. The kinematic
singularities originate on account of non-analyticity of the
phase-volume factor. The non-analyticity, in turn, gives rise to
nonintegrable singularities in the cross-section in the presence
of the MPT-expanded propagators squared. Consequently the latter
singularities is an essential problem in the MPT approach.

In this paper we find a solution to the problem of kinematic
singularities in an arbitrary order of the MPT expansion. The
Coulomb singularities are taken into consideration rather
qualitatively, in the spirit of \cite{Fadin2}--\cite{Fadin4},
which is sufficient for our purposes. We emphasize that we search
for a principle solution to the problem of curing non-integrable
contributions in the cross-section. Correspondingly, we carry out
only analytical calculations necessary for reducing further
calculations to the realizable ones by means of numerical methods.
In doing so we mean that the conventional regular contributions to
the cross-section can be independently calculated in the framework
of conventional PT.

The paper is organized as follows. In Sect.$\,$\ref{not} we
formulate the problem and do its preparatory analysis. An
algorithm of the calculation of the MPT expansion in the case of
the hard-scattering cross-section, is worked out in
Sect.$\,$\ref{pap}. In Sect.$\,$\ref{cross} we calculate singular
contributions to the observable cross-section and introduce a
necessary modification to the method in the near-threshold
regions. In Sect.$\,$\ref{NNLO} we discuss the peculiarities of
the calculations within the NNLO approximation.
Sect.$\,$\ref{disc} outlines the results. In Appendix~A we derive
a formula for the characteristic change of variables in the
$\delta$-function with derivatives. In Appendix~B we carry out
analytical calculations of basic singular integrals.

\section{Statement of the problem}\label{not}
\setcounter{equation}{0}
\def\theequation{\ref{not}.\arabic{equation}}

We analyse pair production and subsequent decay of unstable
particles, in particular in the process of $e^+ e^-$ annihilation.
For simplicity we consider the case when only stable particles are
immediately produced at the decays of the unstable particles. (The
cascade decays should not lead to new significant difficulties.)
The observable cross-section of the whole process is determined
via the convolution of the hard-scattering cross-section with the
flux function describing contributions of nonregistered photons
emitted in the initial state \cite{LEP2}:
\begin{equation}\label{not1}
\sigma (s) = \int\limits_{s_{\mbox{\tiny min}}}^s \frac{\d s'}{s}
\: \phi(s'/s;s) \> \hat\sigma(s')\,.
\end{equation}
Here $s$ is the energy squared in the center-of-mass system,
$\sigma(s)$ and $\hat\sigma(s)$ are the observable and the
hard-scattering cross-sections, respectively. Apart from $s$ both
cross-sections can depend on spin and angular variables. (We do
not consider explicitly this option as the corresponding
modifications are unessential for our analysis.) The ratio
$z=s'/s$ characterizes a fraction of the energy expended on the
production of unstable particles. For our purposes it is
sufficient to take flux function $\phi(z;s)$ in the leading
logarithm approximation. So we put
\begin{equation}\label{not2}
\phi (z;s) = \beta_e (1-z)^{(\beta_e -1)}
             - \frac{1}{2} \beta_e (1+z), \qquad
\beta_e = \frac{2 \alpha }{\pi}\left(\ln \frac{s}{m_e^2}
-1\right).
\end{equation}
The hard-scattering cross-section $\hat\sigma(s)$ must be
determined in the PT framework, though in some cases indirectly
because of divergences arising in the phase-space integral. To
avoid the divergences, usually the Dyson resummation is applied in
the unstable-particles propagators. Starting conceptually from an
exact solution, we initially imply the application of the
Dyson-resummation, too. However further we will asymptotically
expand the propagators squared in the sense of distributions.

Generally, there are two types of contributions to the
hard-scattering cross-section, the factorizable and
non-factorizable ones. The factorizable contributions retain the
structure of the process as the sequential production and decay of
unstable particles. The non-factorizable contributions connect
these subprocesses. Typically the non-factorizable contributions
are represented by configurations with contributions of one of the
unstable particles or of both of them being components of
one-particle irreducible subdiagrams. Such contributions do not
generate singularities in the amplitude and therefore are
integrable in the cross-section. Generally, the non-factorizable
contributions can be related to the cases of inclusive single
production and decay of unstable particles or direct production of
final states. The corresponding calculations in the MPT framework
are simpler as compared to the cases with factorizable
contributions. For this reason we postpone the consideration of
non-factorizable contributions to the end of the next section. The
only exception we make is for soft massless-particles
contributions. The matter is that they can retain the
double-resonant structure even if they formally are
non-factorizable \cite{soft1}--\cite{soft3}. (Actually this
property is a manifestation of general theorem
\cite{Weinberg,Akhiezer} which states that the soft
massless-particles contributions are collected in a factor in the
cross-section.) For this reason we will consider the soft
massless-particles contributions in parallel with the
consideration of factorizable contributions.

So, at first we consider the piece of the cross-section that
possesses the double-resonant structure. The hard-scattering
cross-section in this case is conveniently written as follows:
\begin{equation}\label{not3}
\hat\sigma (s) =
 \int\limits_{\quad {\displaystyle\mbox{\scriptsize $s$}}_
                    { 1 \mbox{\tiny min} } \atop
              \quad {\displaystyle\mbox{\scriptsize $s$}}_
                    { 2 \mbox{\tiny min} }}
            ^{\!\!\infty}   \!\!\!\!\!\!\!\!
 \int\limits^{\infty} \d s_1 \, \d s_2 \;
 \theta(\sqrt{s}-\!\sqrt{s_1}-\!\sqrt{s_2}\,) \,
 \hat\sigma(s\,;s_1,s_2)
 \left(1\!+\!\delta_{c}\right)
\end{equation}
Here $\hat\sigma(s\,;s_1,s_2)$ is an exclusive cross-section, the
$(1+\delta_{c})$ stands for contributions of soft massless
particles, $s_1$ and $s_2$ are the virtualities of the unstable
particles. The $s_{1\,\mbox{\scriptsize min}}$ and
$s_{2\,\mbox{\scriptsize min}}$ are the minimums of $s_1$ and
$s_2$, respectively, defined as squared sums of the masses of the
decay products of the unstable particles. Hereinafter we imply
that the square roots of $s_{1\,\mbox{\scriptsize min}}$ and
$s_{2\,\mbox{\scriptsize min}}$ are less than the masses of the
corresponding unstable particles. The exclusive cross-section
$\hat\sigma(s\,;s_1,s_2)$, we write in the form with extracted
Breit-Wigner (BW) factors:
\begin{equation}\label{not4}
\hat\sigma(s\,;s_1,s_2) =
      \frac{1}{s^2}\sqrt{\lambda (s,s_{1},s_{2})}\;\Phi(s;s_1,s_2)
      \> \rho_{1} (s_{1}) \> \rho_{2} (s_{2})\,.
\vspace*{0.4\baselineskip}
\end{equation}
Here $\rho_{i}(s_{i})$ are the BW factors, $\lambda
(s,s_{1},s_{2}) = [s \!-\!(\sqrt{s_1} \!+\! \sqrt{s_2}\,)^2] [s
\!-\!(\sqrt{s_1} \!- \! \sqrt{s_2}\,)^2]$ is the so-called
kinematic function. The product of the $\sqrt{\lambda
(s,s_{1},s_{2})}$ with the $\theta$-function in (\ref{not3})
constitutes the kinematic factor. Let us note at once that the
kinematic factor has a singularity at $\sqrt{s} =
\sqrt{s_1}+\!\sqrt{s_2}$. The function $\Phi(s;s_1,s_2)$ is the
rest of the amplitude squared. By construction it does not include
singularities associated with the production and decay of unstable
particles. Correspondingly, we consider $\Phi$ determined in the
framework of conventional PT. The BW factors, we define as
follows:
\begin{equation}\label{not5}
\rho_{i} (s_{i}) = \frac{M_{i} \Gamma_{i}}{\pi}\times
|\Delta_{i}(s_{i})|^2\,.
\end{equation}
Here $M_{i}$ is the renormalized mass and $\Gamma_{i}$ is the Born
width of the $i$th unstable particle, $\Delta_{i}$ is a scalar
part of the Dyson-resummed propagator (so the spin factors are
referred to $\Phi$). Further we omit index ``$i\,$'' if its
presence is not necessary. The $\Delta(s)$ has the form
\begin{equation}\label{not6}
\Delta(s) =
 \frac{1}{s - M^2 + \Re \Sigma(s) + \i\:\Im \Sigma(s)}\,,
\end{equation}
where $\Re\Sigma(s)$ and $\Im\Sigma(s)$ are the real and imaginary
parts of the renormalized self-energy. We assume that $\Sigma(s)$
is determined in the on-mass-shell (OMS) scheme of the UV
renormalization, or rather in some version of its generalization
to the unstable-particle case \cite{OMS1}--\cite{Sirlin}. Recall
that the naive expansion of $\Delta(s)$ in powers of the coupling
leads to divergences in integral (\ref{not3}).

Now we proceed directly to arguing the MPT approach. Its most
important feature is the assignment of the distribution sense to
the factors that explicitly or potentially include singularities.
In the given case such factors are the BW factors, the kinematic
factor, and the factor $1+\delta_c$. Function $\Phi$ may be
considered as a representative of trial functions since $\Phi$ is
smooth in the vicinity of prospective singularities (located
actually on the thresholds). By the next step, it is necessary to
asymptotically expand the product of the mentioned factors in the
sense of distributions.

For methodological reasons at first we consider the asymptotic
expansion of the BW factors by considering each of them isolated.
A general structure of their expansion is as follows \cite{F1}:
\begin{equation}\label{not7}
\rho (s) = \delta(s\!-\!M^2) +
 PV \, {\cal T}_N \left\{\,\rho (s)\,\right\} +
 \sum\limits_{n\,=\,0}^N c_{n}(\alpha)\,
 \delta_{n}(s\!-\!M^2) + O(\alpha^{N+1})\,.
\end{equation}
Here the expansion is carried out in powers of $\alpha$ up to
neglected terms of order $O(\alpha^{N+1})$. The leading term
originates by virtue of the relation $\alpha / \left (x^2 +
\alpha^2\right) \to \pi \,\delta(x)$ as $\alpha \to 0$ and the
unitarity relation $\Im \Sigma (M^2) |_{\mbox {\scriptsize
1-loop}} \!= M \Gamma |_{\mbox{\scriptsize Born}}$. The second
term denotes the Taylor (${\cal T}$) expansion in powers of
$\alpha$ up to $\alpha^N$ with the poles in $s-M^2$ defined in the
sense of principal value ($PV$). Recall that the principal value
may be defined by the relation
\begin{equation}\label{not8}
PV \frac{1}{x^n} = \frac{(-)^{n\!-\!1}}{(n\!-\!1)!}\; \frac{\d
^n}{\d x^n} \ln|x|\,,
\end{equation}
where the derivatives are understood in the distributions sense,
i.e.~at an integration they should be switched to the weight
function via formal integration by parts. The sum in (\ref{not7})
corrects the contributions of the $PV$-poles in the points where
the $PV$-prescription operates. The $\delta_{n}(\dots)$ denotes
the $n$th derivative of the $\delta$-function taken with a typical
coefficient,
\begin{equation}\label{not9}
\delta_{n}(x) \equiv \frac{(-)^{n}}{n!}\,\delta^{(n)}(x)\,.
\end{equation}
The highest degree of the derivatives of the $\delta$-functions in
(\ref{not7}) is equal to the highest degree minus 1 of the
$PV$-poles in the Taylor expansion. The coefficients
$c_{n}(\alpha)$ are defined so that to provide the asymptotic
property for the expansion. In this way the contributions of
$\delta$-functions in (\ref{not7}) mean finite counterterms. As
for $c_{n}(\alpha)$ themselves, they are the polynomials in
$\alpha$ with the exponents ranging from $n$ to $N$. The
coefficients of the polynomials are defined by self-energy of the
unstable particle and by its derivatives calculated on the mass
shell. (The contributions of soft massless particles are
considered regularized at this stage.)

For foreseeable applications, seemingly, the expansion within the
NNLO approximation will be sufficient. With this precision formula
(\ref{not7}) takes the form
\begin{eqnarray}\label{not10}
&{\displaystyle \rho(s) \;=\; \delta(s\!-\!M^2) \,+\,
 \frac{M \Gamma_{0}}{\pi}\left\{PV \frac{1}{(s-M^2)^2} -
 PV\,\frac{2\alpha\,\Re\Sigma_1(s)}{(s\!-\!M^2)^3}\right\}}&
 \\
&\displaystyle +\;\sum\limits_{n\,=\,0}^2 c_{n}(\alpha)\,
 \delta_{n}(s\!-\!M^2) + O(\alpha^3)\,.&\nonumber
\end{eqnarray}
Here $\Sigma_1$ is the one-loop self-energy. The $k$-loop
self-energy $\Sigma_k$ is defined through the formula
\begin{equation}\label{not11}
\Sigma(s) = \alpha \,\Sigma_1(s) + \alpha^2 \,\Sigma_2(s) +
\alpha^3 \,\Sigma_3(s) + \cdots.
\end{equation}
The polynomials $c_{n}(\alpha)$ within the NNLO approximation
actually have been calculated in \cite{F1}, though have been
presented not quite explicitly. Below we represent $c_{n}(\alpha)$
in completely expanded form in an arbitrary version of the above
mentioned generalization of the OMS scheme. All these versions are
characterized by the one-loop conditions $\Re\Sigma_{1}(M^2)=0$,
$\Re\Sigma_{1}^{\,\prime}(M^2)=0$. So we have
\begin{eqnarray}\label{not12}
c_0 & = & - \, \alpha \, \frac{I_2}{I_1} + \alpha^2
\left(\frac{I_2^2}{I^2_1} - \frac{I_3}{I_1} - \frac{1}{2}\,I_1
I_{1}^{\,\prime\prime} + R_2 \frac{I_1^{\,\prime}}{I_1}-
R_2^{\,\prime}\right),
\\[0.5\baselineskip]
c_1 &=& -\, \alpha^2 \left(I_1 I_1^{\,\prime} + R_2\right), \qquad
c_2 \;\;=\; -\, \alpha^2 I^2_1 \,.\nonumber
\end{eqnarray}
Here $I_{k} = \Im\,\Sigma_{k}(M^2)$, $R_{k} =
\Re\,\Sigma_{k}(M^2)$, and the primes mean the derivatives at
$s=M^2$. The $R_2$ and $R_2^{\,\prime}$ are determined by the
renormalization conditions, too, though differently in different
versions of the generalization of the OMS scheme \cite{OMS-bar}.
In particular, in the $\overline{\mbox{OMS}}$ or ``pole'' scheme
\cite{OMS-bar,Sirlin}, $R_2 = - I_1 I_{1}^{\,\prime}$ and
$R_2^{\,\prime} = - I_1 I_{1}^{\,\prime\prime}/2$.

So, we have determined the asymptotic expansion of isolated BW
factors. Now we consider the product of BW factors. It is clear
that in the case of smooth weight the asymptotic expansion of the
product has the form of the formal product of the expansions of
separately taken BW factors. However, this is not the case if the
weight includes singularities which intersect with the
singularities of the expanded BW factors. Unfortunately, this
occurs in the case of integral (\ref{not3}) because of the soft
massless-particles contributions and the singularity of the
kinematic factor. In order to asymptotically expand, nevertheless,
the integrand in (\ref{not3}), we should independently make
definition of the expansion of the product of BW factors in the
vicinities of intersections of singularities.

At first we discuss the case of the singularities caused by the
soft massless-particles contributions. Recall that we symbolically
expressed them through the factor $1+\delta_c$ in formula
(\ref{not3}). Actually these contributions become crucial when the
soft massless-particles momenta merge with the momenta of the
unstable particles (owing to the emission or absorbtion of
massless particles) and when the unstable-particles momenta turn
out to be on the mass shell. The asymptotic expansion in this case
can be fulfilled if the additional singularities, intersecting
with the singularities of the BW factor, are cancelled in the
cross-section. In reality the cancellation takes place in the case
of the ordinary IR-divergent contributions, but not in the case of
Coulomb singularities. Therefore a solution to the problem of soft
massless-particles contributions can exist in the case only when
Coulomb singularities are somehow regularized.

Let us postpone the problem of the Coulomb singularities and
concentrate on the ordinary IR-divergent contributions. A solution
to this problem actually is proposed in \cite{F1}. It is based on
the introduction of dimension regularization for the IR
singularities and then on the calculation of additional
counterterms at the intersections of singularities. Apart from the
$\delta$-functions of virtualities of the unstable particles,
these additional counterterms include also the $\delta$-functions
of the massless-particles momenta. Moreover, the coefficients at
these counterterms are found to be singular in the
dimension-regularization parameter. However the later
singularities are to be cancelled after the integration over the
massless-particles momenta.\footnote{The basic part of the
calculation of the additional counterterms in the case of pair
production and decay of unstable particles, at the level of the
one-photon contributions, has been made by the author of this
article in collaboration with F.Tkachov \cite{F2}.} In this way
the counterterms become identical in form with those in the case
without the contributions of soft massless particles (though they
appear with modified coefficients). An alternative solution is
based on the introduction of the IR regularization through
inserting of a technical mass for massless particles. A remarkable
property of this regularization is the absence in its framework of
the above mentioned intersections of singularities \cite{N1}. By
virtue of this fact the extra counterterms do not arise at all in
this case. So, again, after the integration over the
massless-particles momenta the structure of the expansion becomes
the same as in the case without the contributions of massless
particles. For this reason we further do not consider explicitly
the contributions of soft massless particles except contributions
that lead to Coulomb singularities.

Proceeding to the Coulomb singularities, we note first of all that
they arise in a more restricted kinematic region and have
different nature as compared to the ordinary IR-divergent
contributions. So from the very beginning they can be isolated
from the ordinary soft massless-particles contributions and
thereupon can be separately investigated. By this means it was
found that the Coulomb singularities arise on the mass shell and
simultaneously in the limit of small relative velocities of
outgoing massive particles. At the level of the cross-section they
may be collected in the so-called Coulomb factor. For the first
time this factor was discovered in the case of stable particles
and in electrodynamics with the result \cite{Coulomb1,Coulomb2}
(hereinafter we omit the ordinary IR-divergent contributions)
\begin{equation}\label{not13}
1+\delta_{c} = \left|\,\psi(0)\right|^{\,2} =
\frac{X}{1-\exp{(-X)}}\,, \qquad\qquad X =
\frac{\kappa\,\alpha}{\beta_0}\,.
\end{equation}
Here $\psi(0)$ is the wave function at the origin of the charged
massive particles moving in the center-of-mass frame (c.m.f.) with
velocity $\beta_0$, $\kappa$ is a positive coefficient. (In the
case with color particles in QCD the Coulomb factor is the same
with the modification in the coefficient $\kappa$ only
\cite{Fadin5,Fadin6}). In the expanded form factor (\ref{not13})
can be represented with the aid of the formula
\begin{equation}\label{not14}
\frac{X}{1-\exp{(-X)}} = 1 +\frac{X}{2}+
\sum\limits_{n=1}^{\infty} B_{2n} \frac{X^{2n}}{(2n)!} \,,
\end{equation}
where $B_n$ are Bernoulli numbers.

Returning to the MPT expansion, we note that the $PV$-poles and
the derivatives of the $\delta$-functions in the expansions of the
BW factors are ill-determined in the presence of the Coulomb
singularities. Really, in accordance with the integration rules of
the $PV$-poles and the derivatives of the $\delta$-functions the
Coulomb factor must be taken into consideration not only on-shell,
but also with the derivatives with respect to virtualities of
unstable particles. However the derivatives of the Coulomb
singularities cannot be determined as they appear precisely on the
mass~shell.

In this connection further we use an extended representation of
the Coulomb factor available in the unstable-particle case. Namely
we mean a particular resummation of the singular and some kind of
nonsingular Coulomb contributions, which should be made at a given
number of exchanges by soft massless particles. The resummation
should be made for the off-shell outgoing unstable particles and,
simultaneously, with the Dyson resummation in those propagators of
the unstable particles that are directly involved in the
generation of the Coulomb singularities.\footnote{The mentioned
propagators belong to one-particle irreducible subdiagrams.
Therefore they by no means are involved in the BW factors that are
to be MPT-expanded.} The practical extraction of the mentioned
contributions is a rather subtle task. For details we refer to
\cite{Fadin2}--\cite{Fadin4}, where the extraction is made both
for one- and multi-photon contributions. Below we adduce the
result with the one-photon contribution. In the notation of
\cite{Bardin}, but omitting some superfluous term \cite{Fadin4}
which is inessential for the validity of the ultimate result, we
have
\begin{equation}\label{not15}
{\delta_c}_{\bigl|\,\mbox{\footnotesize one-photon}} =
\frac{\kappa \alpha}{2\beta}\left[ 1 - \frac{2}{\pi}\,\arctan \!
\left( \frac{|\beta_{M}|^2 - \beta^2}{2\beta \, \Im\beta_{M}}
\right)\right] .
\end{equation}
Here $\beta = s^{-1} \sqrt{\lambda (s,s_{1},s_{2})}$ is the
velocity of outgoing unstable particles in the center-of-mass
frame, $\beta_{M} = \sqrt{1-4 (M^2\!- \i M \Gamma)/s}$, $\Gamma$
is the width of unstable particles. Notice that with nonzero
$\Gamma$ the r.h.s.~in (\ref{not15}) is a smooth function of
$\beta$. At the same time at going to the mass-shell and
subsequently taking the limit $\Gamma \to 0$, the r.h.s.~becomes
the $\kappa \alpha /(2\beta_0)$, which is the one-photon Coulomb
singularity.

We emphasize that in fact only imaginary part of the on-shell
self-energy may be involved in the above resummation. Therefore
the regularized Coulomb factor may be made gauge-invariant in the
$\overline{\mbox{OMS}}$ or ``pole'' scheme of the UV
renormalization \cite{OMS-bar,Sirlin}, because in this scheme both
the renormalized mass and the imaginary part of the on-shell
self-energy are gauge invariant quantities. The gauge invariance
of the regularized Coulomb factor, in turn, implies the gauge
invariance of the piece of the cross-section that remains after
the extraction of the Coulomb factor. So provided this remaining
piece is completely expanded in powers of $\alpha$, the
contributions at each power of $\alpha$ must possess the property
of gauge cancellations. In the next section we show the existence
of the complete expansion of the above mentioned remaining piece
of the cross-section.

So we have gained the regularity property of the Coulomb factor
and the possibility to differentiate it with respect to the
virtualities of the unstable particles. In addition, we have a
regularity of the derivatives of the Coulomb factor, too. The
latter property most easily may be shown by noting that the
regularized Coulomb factor is an even function of $\beta$. In the
multi-photon case this is evident from the appropriate explicit
expression \cite{Fadin4}. In the one-photon case this becomes
clear if proceeding to the representation
\begin{equation}\label{not16}
{\delta_c}_{\bigl|\,\mbox{\footnotesize one-photon}} =
\frac{\kappa \alpha}{\pi\beta}\,\arctan \! \left( \frac{2\beta \,
\Im\beta_{M}}{|\beta_{M}|^2 - \beta^2} \right) ,
\end{equation}
which is equivalent to (\ref{not15}) at small enough $\beta$. From
the even property it follows that $\delta_c$ depends effectively
on $\beta^2$. (In particular this is evident from the Taylor
expansion.) This implies that both the $\delta_c$ and the
derivatives of $\delta_c$ with respect to $s_i$ are smooth
functions of $s_i$.

The property of smoothness of the regularized Coulomb factor and
also of its derivatives with respect to the virtualities allows us
to refer the Coulomb factor to the weight in formula (\ref{not3}).
Correspondingly, at the further calculations we consider factor
$1+\delta_c$ as absorbed by function $\Phi$. In doing so, the
$1+\delta_c$ may be taken into consideration in the expanded form
with respect to the number of exchanges by soft massless
particles. Unfortunately, this is not the complete expansion since
some dependence on $\alpha$ is involved in the parameter of the
regularization. Nevertheless with the fixed parameter $\Gamma$,
the expansion is an asymptotic one with respect to $\alpha$.

Finally, we turn to the kinematic factor
$\theta(\!\sqrt{s}-\!\sqrt{s_1}-\!\sqrt{s_2}\,)
\sqrt{\lambda(s,s_{1},s_{2})}$. As noted above, its singularity
located on the point set $\sqrt{s} - \!\sqrt {s_1} -
\!\sqrt{s_2}=0$ can intersect with the singularities of the
expanded BW factors. Most easily this may be shown by imposing
conditions on $s$. So, at $\sqrt{s} = \sqrt{s_j} + M_i$ ($i\not =
j$) the point set $\sqrt{s} - \!\sqrt {s_1} - \!\sqrt{s_2}=0$ is
reduced to the isolated point $\sqrt{s_i} = M_i$, where the
singularity of one of the expanded BW factors is located. At
$\sqrt {s} = M_1 + M_2$ the above point set is reduced to
$\sqrt{s_1} + \! \sqrt{s_2} = M_1 + M_2$, where the singularities
of both BW factors are simultaneously located.

At first sight this property makes senseless the rules of the
integration of the $PV$-poles and of the derivatives of the
$\delta$-functions, and hence the very possibility of the
calculation of the asymptotic expansion of integral (\ref{not3})
becomes questionable. However, the distribution method in this
case again allows us to add a sense to the integrals that arise in
the course of the expansion. Moreover, the ambiguities which are
peculiar to this method may be again removed by means of the
imposing of the asymptotic condition of the expansion.
Furthermore, both these requirements are automatically satisfied
by application of the analytical regularization of the kinematic
factor. In the next section we proceed to a systematic discussion
of this~issue.

\section{The scheme of calculation of
{\large $\hat{\sigma}(s)$}}\label{pap} \setcounter{equation}{0}
\def\theequation{\ref{pap}.\arabic{equation}}

Fist of all we reformulate the problem in dimensionless variables.
For this purpose we introduce dimensionless energy variables $x$,
$x_1$, $x_2$ counted off from thresholds,
\begin{equation}\label{pap1}
\sqrt{s} = 2 M + \frac{M}{2} \, x\,, \quad \sqrt{s_{i}} = M_i +
\frac{M}{2} \, x_{i}\,.
\end{equation}
Here $i = 1,2$, and $M = (M_1 + M_2)/2$. Then formula (\ref{not3})
takes the form
\begin{eqnarray}\label{pap2}
\widetilde{\sigma}(x) &=& \int\!\!\!\int\!\d x_1 \, \d x_2
\;\theta(x_{1}\!+\!a_{1}) \theta(x_{2}\!+\!a_{2})
\theta(x\!-\!x_1\!-\!x_2) \sqrt{x\!-\!x_1\!-\!x_2} \\
&& \qquad \times \> \widetilde{\Phi}(x\,;x_1,x_2)
\left(1+\widetilde{\delta}_{c}\right) \widetilde{\rho}_1(x_1)
\widetilde{\rho}_2(x_2)\,.\nonumber
\end{eqnarray}
Here $\widetilde{\sigma}(x) = M_{1} M_{2} \> \hat\sigma (s)$,
$\widetilde{\rho}_i(x_i) = M M_{i} \, \rho_i(s_i)$, and $a_{i} =
2(M_i - \sqrt{s_{i\mbox{\tiny min}}})/M$. At once we notice that
the parameters $a_{i}$ are strictly positive at
$M_i^2>s_{i\mbox{\tiny min}}$ and that in any case $a_1\!+\!a_2 <
4$. Notice also that in view of the $\theta$-functions, the
$\widetilde{\sigma}(x)$ is effectively proportional to
$\theta(x\!+\!a_1\!+\!a_2)$. The $1+\widetilde{\delta}_{c}$ in
(\ref{pap2}) is equivalent to within the change of variables to
$1+\delta_{c}$ in (\ref{not3}). Functions $\widetilde{\Phi}$ and
$\Phi$ are related each other by the formula:
\begin{equation}\label{pap3}
\d x_1\,\d x_2\>\sqrt{x\!-\!x_1\!-\!x_2} \;
\widetilde{\Phi}(x\,;x_1,x_2) = \frac{1}{M^2} \: \d s_1 \, \d s_2
\, \frac{1}{s^2}\,\sqrt{\lambda(s;s_{1},s_{2})} \;
\Phi(s;s_1,s_2)\,.
\end{equation}
Since the factors $\sqrt{x\!-\!x_1\!-\!x_2}$ and
$\sqrt{\lambda(s;s_{1},s_{2})}$ have identical analytical
properties in the integration area, the analytical properties of
$\widetilde{\Phi}$ and $\Phi$ are identical in this area, too. Let
us remember that by the construction $\Phi$ is smooth in the
vicinity of physical thresholds associated with the production and
decay of unstable particles. So $\Phi(s;s_1,s_2)$ has no
singularities in the points $\sqrt{s}=2M$, $\sqrt{s_{i}}=M_i$,
$\sqrt{s}-\sqrt{s_{1\,\mbox{\scriptsize
min}}}-\sqrt{s_{2\,\mbox{\scriptsize min}}}=0$,
$\sqrt{s}-\sqrt{s_{i\,\mbox{\scriptsize min}}}-M_j=0$,
$\sqrt{s}-\sqrt{s_{i\,\mbox{\scriptsize min}}}-\sqrt{s_j}=0$,
$\sqrt{s}-\sqrt{s_i}-M_j=0$,
$\sqrt{s}-\sqrt{s_{1}}-\sqrt{s_{2}}=0$ ($i \not= j$).
Correspondingly, $\widetilde{\Phi}(x\,;x_1,x_2)$ has no
singularities in the points $x=0$, $x_i=0$, $x+a=0$, $x+a_i=0$,
$x+a_i-x_j=0$, $x-x_i=0$, $x\!-\!x_1\!-\!x_2=0$.

The asymptotic expansion of isolated BW factors
$\widetilde{\rho}_i(x_i)$ most easily may be obtained by means of
change of variables in formula (\ref{not7}). So, let us rewrite
the second relation in (\ref{pap1}) in the form $s_i-M_i^2 = M M_i
\,x_i [1 + x_i \, M/(4 M_i)]$, and note that $[1 + x_i \, M/(4
M_i)] \not=0$ everywhere in the integration area. This implies
that the singular point $s_i-M_i^2 = 0$ transforms into the point
$x_i = 0$ in new variables. So the $\delta(s_i \!- \! M_i^2)$
becomes the $(MM_i)^{-1}\delta(x_i)$, while the derivatives of
$\delta(s_i \!- \! M_i^2)$ transforms into the sum of
$\delta(x_i)$ and its derivatives. The final formula valid within
the integration area is derived in Appendix~A,
\begin{equation}\label{pap4}
\delta_{n} (s_i-M_i^2) = \frac{1}{\left(M M_i\right)^{n+1}} \,
\sum_{k=0}^{n} {n+k \choose n} \!\! \left(-\frac{M}{4
M_i}\right)^{\!k} \delta_{n-k}(x_i)\,.
\end{equation}
Recall that the $\delta_{n}$ means the $n$-th derivative of the
$\delta$-function with a typical coefficient, cf.~(\ref{not9}). As
long as the $PV$-regularization is a canonical one \cite{Gelfand},
the change of variables in the $PV$-poles is made as in
conventional functions, i.e.~without the adding of a sum of the
$\delta$-function and its derivatives:
\begin{equation}\label{pap5}
PV \frac{1}{\left(s_i-M_i^2\right)^n} = \frac{1}{\left(M
M_i\right)^n} \left(1+\frac{M}{4 M_i} x_i \right)^{\!\!-n}PV
\frac{1}{x_i^n}\:.\quad
\end{equation}
Substituting (\ref{pap4}) and (\ref{pap5}) into (\ref{not7}), we
get the asymptotic expansion of $\widetilde{\rho}_i (x_i)$.

Unfortunately, the presence of the kinematic factor
$\theta(x\!-\!x_1\!-\!x_2) \sqrt{x\!-\!x_1\!-\!x_2}$ does not
allow us to substitute the expansions for $\widetilde{\rho}_i
(x_i)$ in formula (\ref{pap2}), because the derivatives of the
kinematic factor are non-integrable, or even completely
undetermined in the point $\{ x\!-\!x_i\! = 0,\,x_j\! = 0 \}$, $i
\not= j$. Nevertheless the situation changes if to proceed to the
analytically regularized kinematic factor, namely to
$\theta(x\!-\!x_1\!-\!x_2)\,(x\!-\!x_1\!-\!x_2)^{\lambda}$ with
non-integer $\lambda$.\footnote{Let us remark that
$(x\!-\!x_1\!-\!x_2)^{\lambda}$ instead of
$\sqrt{x\!-\!x_1\!-\!x_2}$ arises automatically if the
phase-volume integration is carried out at the space dimension $d
= 4+2\varepsilon$, $\varepsilon = \lambda-1/2$.} With large enough
$\lambda$ this gives a regularity property for the derivatives. As
a result, after the substitution of the expansions for
$\widetilde{\rho}_i (x_i)$, the integral (\ref{pap2}) considered
as iterated becomes calculable. Further, after the integral is
calculated, we have to take the limit $\lambda \to 1/2$ in order
to match the original integral. So the problem is reduced to the
taking of the limit of the already calculated integral. Notice
that the above procedure means the treatment of the kinematic
factor as the distribution $(x\!-\!x_1\!-\!x_2)_{+}^{1/2}$
\cite{Gelfand}.

Fortunately, if the above mentioned limit exists then this
guarantees the asymptotic property of the expansion of the whole
of integral (\ref{pap2}). This follows, first of all, from the
fact that prior to any expansion the substitution of the
distribution $(x\!-\!x_1\!-\!x_2)_{+}^{1/2}$ for the kinematic
factor does not mean a modification of the integral. This, in
particular, means that the asymptotic expansion after the
substitution is a completely rightful operation. Technically the
expansion may be carried out through the proceeding at
intermediate stages of calculations to the large enough~$\lambda$.
(The infimum of $\lambda$ is dependent on the order up to which
the expansion is to be made.) This allows us to treat the
regularized kinematic factor as a smooth weight and on this basis
to asymptotically expand the BW factors. The integrals that will
appear after the expansion by our construction will be calculable.
Further, the analyticity will allow us to vary $\lambda$ with
keeping the asymptotic property of the expansion. The existence of
the limit at each term of the expansion will imply the
implementation of asymptotic property of the expansion.

Now let us proceed to the calculation of the above mentioned
integrals at arbitrary but large enough $\lambda$. For this
purpose, we consider any one term of the resulting expansion of
the integrand in (\ref{pap2}) that appears after the formal
expansion of the BW factors. Let this term include $PV x_1^{-n_1}$
or $\delta_{n_1\!-\!1}(x_1)$ from one of the BW factors and $PV
x_2^{-n_2}$ or $\delta_{n_2\!-\!1}(x_2)$ from another BW factor
($n_{1,2} \ge 1$). Recall that we consider the kinematic factor in
the form $(x\!-\!x_1\!-\!x_2)_{+}^{\lambda}$ having in mind that
at the intermediate stage the $\lambda$ is large enough. The
$\theta(x_1 + a_1) \, \theta(x_2 + a_2)$, we consider as a
potentially singular factor, as these $\theta$-functions via the
integration by parts can generate non-integrable singularities at
$\lambda = 1/2$. As a pure weight we consider the function
$\widetilde{\Phi}(x\,;x_1,x_2)$ multiplied by the regularized
Coulomb factor and by the definite regular coefficients that arise
in the course of application of the formulas (\ref{not7}),
(\ref{pap4}) and (\ref{pap5}). For brevity we denote the mentioned
weight by the same symbol $\widetilde{\Phi}(x\,;x_1,x_2)$.

Further, let us subtract and add to $\widetilde{\Phi}$ the $n_1
\times n_2$ of the first terms of its Taylor expansion in powers
of $x_1$ and $x_2$,
\begin{equation}\label{pap6}
{\cal T}_{n_1\!,n_2} \!\! \left\{
{\widetilde{\Phi}(x\,;x_1,x_2)}\right\} = \!\!\sum_{k_1=0}^{n_1-1}
\sum_{k_2=0}^{n_2-1}  \frac{x_1^{k_1}}{k_1 !}\,
\frac{x_2^{k_2}}{k_2 !} \:
\widetilde{\Phi}^{(k_1,\,k_2)}(x\,;0,0).
\end{equation}
Then $\widetilde{\Phi}$ takes the form of a sum of two terms, the
first one is (\ref{pap6}) and the second one is the difference
$\Delta\widetilde{\Phi} = \widetilde{\Phi} - {\cal
T}\{\widetilde{\Phi}\}$. The difference, we represent in the form
\begin{eqnarray}\label{pap7}
&{\Delta \widetilde{\Phi}(x\,;x_1,x_2) = {\displaystyle
\sum_{k_1=0}^{n_1-1} \; \frac{x_1^{k_1}}{k_1 !} \; \Bigl[
x_2^{n_2} \, \varphi_{n_2}^{(k_1)}(x,x_2)\Bigr] +
\sum_{k_2=0}^{n_2-1} \; \frac{x_2^{k_2}}{k_2 !} \; \Bigl[
x_1^{n_1} \, \varphi_{n_1}^{(k_2)}(x,x_1)\Bigr]}}&
 \nonumber\\[0.5\baselineskip]
&  + \; x_1^{n_1} x_2^{n_2} \, \varphi(x\,;x_1,x_2)\,.&
\end{eqnarray}
Here the first sum represents the Taylor expansion up to
$x_1^{n_1-1}$ of the remainder of the Taylor expansion of
$\widetilde{\Phi}(x\,;x_1,x_2)$ in $x_2$ up to $x_2^{n_2-1}$. The
second sum has the same meaning with the replacement $x_1
\leftrightarrow x_2$. The third term has the meaning of the common
remainder. Note that by the construction the functions
$\varphi_{n_j}^{(k_i)}(x,x_j)$ and $\varphi(x\,;x_1,x_2)$ are
regular in some neighborhoods of the points $x_1 = 0$, $x_2 = 0$.

It is clear that ${\cal T}\{\widetilde{\Phi}\}$ leads to singular
contributions in the presence of the $PV$-poles and the
$\delta$-functions. However, thanks to its power-like dependence
on $x_1$ and $x_2$ the integral (\ref{pap2}) of these
contributions can be analytically calculated. Namely, owing to the
relations
\begin{eqnarray}\label{pap8}
 (y-x)_{+}^{\lambda}\,x^k\,\delta_{n}(x) & \;=\; &
 (y-x)_{+}^{\lambda}\,\delta_{n-k}(x)\,,\qquad \;\; 0 \le k \le n
 \,,
 \\[0.3\baselineskip]
 \label{pap9}
 (y-x)_{+}^{\lambda}\,x^k \; PV \frac{1}{x^{n}} & \;=\; &
 (y-x)_{+}^{\lambda}\,PV \frac{1}{x^{n-k}}\,, \qquad 0 \le k < n
 \,,
\end{eqnarray}
which are valid in the framework of the analytical regularization,
the mentioned contributions to (\ref{pap2}) are reduced to linear
combinations of the basic integrals
\begin{eqnarray}\label{pap10}
\!\!\!\!\!\!\!\!\!\!\!\!\!\!\!A^{\,\lambda}_{k_1 k_2}(x) &=&
 \int\!\!\!\int\!\d x_1 \, \d x_2 \;
 \theta(x_{1}\!+\!a_{1})\theta(x_{2}\!+\!a_{2})\,
 (x\!-\!x_1\!-\!x_2)^{\lambda}_{+}\,
 \delta_{k_1\!-\!1}(x_1)\,\delta_{k_2\!-\!1}(x_2)\,,
 \\[0.5\baselineskip]\label{pap11}
\!\!\!\!\!\!\!\!\!\!\!\!\!\!\!B^{\,\lambda}_{k_1 k_2}(x) &=&
 \int\!\!\!\int\!\d x_1 \, \d x_2 \;
 \theta(x_{1}\!+\!a_{1})\theta(x_{2}\!+\!a_{2})\,
 (x\!-\!x_1\!-\!x_2)^{\lambda}_{+}\,
 PV \frac{1}{x_1^{k_1}}\,\delta_{k_2\!-\!1}(x_2)\,,
 \\[0.5\baselineskip]\label{pap12}
\!\!\!\!\!\!\!\!\!\!\!\!\!\!\!C^{\,\lambda}_{k_1 k_2}(x) &=&
 \int\!\!\!\int\!\d x_1 \, \d x_2 \;
 \theta(x_{1}\!+\!a_{1})\theta(x_{2}\!+\!a_{2})\,
 (x\!-\!x_1\!-\!x_2)^{\lambda}_{+}\,
 PV \frac{1}{x_1^{k_1}}\,PV \frac{1}{x_2^{k_2}}\,.
\end{eqnarray}
Here subscripts $k_i$ range over $1 \le k_i \le n_i$. Notice that
among the $B$-type integrals that include both the $PV$-pole and
the $\delta$-function, there are integrals with the opposite order
of $PV$ and $\delta$. Let us denote these integrals by
$\overline{B\,}$$^{\,\lambda}_{k_1 k_2}$. We do not write down
them explicitly as they may be obtained immediately by
substitutions $\{k_1,a_1\} \leftrightarrow \{k_2,a_2\}$ in formula
(\ref{pap11}).

The analytical calculation of integrals
(\ref{pap10})--(\ref{pap12}) is carried out in Appendix~B. In the
general case the outcomes have the form of a sum of regular and
singular at $\lambda = 1/2$ contributions. The singular
contributions are defined as the power-like distributions of the
type $x_{+}^{\lambda}$ multiplied by some regular factors, see
Table~\ref{table1}.
\begin{table}[t]
\caption{The singular contributions to $\widetilde{\sigma}(x)$
arising via the two-fold basic integrals ($r$ is a non-negative
integer such that $5/2\!-n_2\!+r<0$).} {\begin{tabular}{l l l}
\\[-2mm]\hline\\[-2mm]
 $\qquad\quad A^{1/2}_{n_1 n_2}$ $\qquad\qquad\qquad\quad\;$
          & $x_{+}^{5/2-n_1\!-n_2}$ &         \\[3mm]
 $\qquad\quad B^{1/2}_{n_1 n_2}$  $\qquad\qquad\qquad\quad\;$
          & $(-x)_{+}^{5/2-n_1\!-n_2}$ $\qquad\qquad\quad\;$
          & $(x+a_1)_{+}^{5/2-n_2+r} \qquad$   \\[3mm]
 $\qquad\quad C^{1/2}_{n_1 n_2}$  $\qquad\qquad\qquad\quad\;$
          & $x_{+}^{5/2-n_1\!-n_2}$ $\;\;\,$
          &       \\[1mm]
\hline
\end{tabular}\label{table1}}
\end{table}
The calculation of the observable cross-section in the presence of
the singular distributions is considered in the next section. Here
we remark only that with a complicated weight the integral of the
distribution $x_{+}^{\lambda}$ cannot be analytically calculated,
but can be calculated numerically with the aid of formula
\cite{Gelfand}
\begin{equation}\label{pap13}
\int \d x \; x_{+}^{\lambda} \,\psi(x) = \int\limits_{0}^{\infty}
\d x \; x^{\lambda} \left\{\psi(x) -
 \sum_{k=0}^{K-1} \frac{x^{k}}{k!} \, \psi^{(k)}(0)\right\}.
\end{equation}
Here $\lambda < -1$ and $\lambda$ is non-integer, $K$ is a
positive integer such that $-K\!-\!1< \lambda < -K$, $\psi(x)$ is
a test function which must be smooth enough in the vicinity of
$x=0$.

In the difference term $\Delta \widetilde{\Phi}(x\,;x_1,x_2)$, we
consider separately the contributions of the sums and of the
common remainder. In the common remainder in view of the relations
\begin{equation}\label{pap14}
x^n\,\delta_{n-1}(x) = 0\,,\qquad x^n \, PV x^{-n} =1\,,
\end{equation}
the multipliers $x_1^{n_1}$ and $x_2^{n_2}$ make vanish
contributions of the $\delta$-functions and cancel the $PV$-poles.
In order to calculate the remaining integral, we proceed to the
cone variables $x_1 + x_2 = \xi$, $x_1 - x_2 = 2 \eta$ and make a
shift $\xi \to x - \xi$. After the calculation of the integral $\d
\eta$, the relevant contribution to (\ref{pap2}) takes the form of
\begin{equation}\label{pap15}
\int \d \xi \>\, \theta(x+a_1+a_2-\xi)\,\xi_{+}^{\lambda}
\,\varphi(x,\xi)\,.
\end{equation}
Here $\varphi(x,\xi)$ is a regular function that arises after the
calculation of the integral $\d \eta$. In the case $\lambda = 1/2$
integral (\ref{pap15}) is absolutely convergent and therefore can
be numerically calculated.

In the sums in $\Delta \widetilde{\Phi}(x\,;x_1,x_2)$, the
multiplier $x_j^{n_j}$ makes vanish contributions of
$\delta_{n_j-1}(x_j)$ and cancels $PV x_j^{-n_j}$. The
contributions of $\delta_{n_i-1}(x_i)$ and $PV x_i^{-n_i}$ remain
not touched upon, but the dependence on $x_i$ is simpler.
Therefore at first we calculate the integral $\d x_i$. In view of
(\ref{pap8}) and (\ref{pap9}), it is reduced to the sum of the
basic integrals ($1 \le k \le n_i$)
\begin{equation}\label{pap16}
{\tt I}^{\lambda}_{k}(x-x_j) = \int \d x_i \;\theta(x_i + a_i) \,
(x-x_i-x_j)_{+}^{\lambda}\,\delta_{k-1}(x_i)\,,\\
\end{equation}
\begin{equation}\label{pap17}
{\tt J}^{\lambda}_{k}(x-x_j) = \int \d x_i \;\theta(x_i + a_i) \,
(x-x_i-x_j)_{+}^{\lambda}\; PV \frac{1}{x_i^k}\,.\quad\!
\end{equation}
The analytical calculation of these integrals is carried out in
Appendix~B, as well. Having analytical expressions for ${\tt
I}^{\lambda}_{k}$ and ${\tt J}^{\lambda}_{k}$, we can calculate
the remaining integral $\d x_j\,$. In the general case it is
reduced to the sum of integrals of the kind
\begin{equation}\label{pap18}
\int \d x_j \;\theta(x_j + a_j) \, \varphi_{n_j}^{(k_i)}(x,x_j) \,
{\tt P}^{\lambda}_{k}(x-x_j)\,.
\end{equation}
Here ${\tt P}^{\lambda}_{k}$ stands for ${\tt I}^{\lambda}_{k}$ or
${\tt J}^{\lambda}_{k}$ and $\varphi_{n_j}^{(k_i)}(x,x_j)$ is
introduced in (\ref{pap7}). By making the change of variable
$x-x_j \to \xi$, we rewrite (\ref{pap18}) in a more convenient
form,
\begin{equation}\label{pap19}
\int \d \xi \;\theta(x + a_j - \xi) \,
\varphi_{n_j}^{(k_i)}(x,x-\xi) \, {\tt P}^{\lambda}_{k}(\xi)\,.
\end{equation}

At first we consider the case ${\tt P}^{\lambda}_{k} = {\tt
I}^{\lambda}_{k}$. In view of ${\tt I}^{\lambda}_{k}(\xi) \sim
\xi_{+}^{1+\lambda-k}$, integral (\ref{pap19}) is absolutely
convergent at $\lambda > k-2$. At $\lambda < k-2$ and $x+a_j \not
= 0$ the integral can be calculated with the aid of (\ref{pap13}).
The condition $x+a_j \not = 0$ is necessary for the providing of
smoothness for the weight at $\xi=0$. As $x+a_j \to 0$, integral
(\ref{pap19}) tends to infinity. Nevertheless, the singular
contribution can be analytically calculated and, simultaneously,
the regular background can be numerically calculated. Namely we
represent $\varphi_{n_j}^{(k_i)}(x,x-\xi)$ in the form of the
Taylor expansion with a remainder,
\begin{equation}\label{pap20}
\varphi_{n_j}^{(k_i)}(x,x-\xi) = \sum\limits_{r=0}^{K-1}
\;\frac{\xi^r}{r!} \; \varphi_{n_j}^{(k_i,\,r)}(x,x) +
\Delta\varphi_{n_j}^{(k_i)}(x,x-\xi)\,.
\end{equation}
Here $-K\!-\!1 < 1+\lambda-k < \!-K$, and
$\varphi_{n_j}^{(k_i,\,r)}(x,x) = \partial^r/\partial\xi^r \,
\varphi_{n_j}^{(k_i)}(x,x\!-\!\xi)|_{\,\xi=0}$. Owing to
$\Delta\varphi_{n_j}^{(k_i)}(x,x\!-\!\xi)=O(\xi^K)$, the integral
with $\Delta\varphi_{n_j}^{(k_i)}$ is absolutely convergent, and
this contribution can be numerically calculated. The integral of
the sum in (\ref{pap20}) gives
\begin{equation}\label{pap21}
\sum\limits_{r=0}^{K-1}
\;\frac{(x+a_j)_{+}^{2+\lambda-k+r}}{2+\lambda-k+r} \;
\frac{\varphi_{n_j}^{(k_i,\,r)}(x,x)}{r!}\,.
\end{equation}
Here $2\!+\!\lambda\!-\!k\!+\!r < 0$, and we note the distribution
sense of the singular contributions. Up to a coefficient, formula
(\ref{pap21}) determines the appropriate singular contributions to
$\widetilde{\sigma}(x)$.

In the case ${\tt P}^{\lambda}_{k} = {\tt J}^{\lambda}_{k}$, we
have generally two singularities in ${\tt P}^{\lambda}_{k}(\xi)$,
the $(-\xi)_{+}^{1+\lambda-k}$ and $\xi_{+}^{1+\lambda-k}$,
cf.~(\ref{calc14}). However, the latter singularity does not make
contributions at $\lambda=1/2$ in view of the factor
$\cos(\pi\lambda)$. To calculate the contribution of the former
singularity, we do the change of variable $\xi \to -\xi$ and take
advantage of the relation $\theta(x+a_j+\xi) = 1 -
\theta(-x-a_j-\xi)$. It is clear that the unity gives regular
contribution. The $\theta(-x-a_j-\xi)$ gives a regular
contribution outside of a neighborhood of $x + a_j = 0$. Inside
the neighborhood, by complete analogy with the previous case, it
gives a sum of the regular contribution and the power-like
singular contributions $(-x - a_j)_{+}^{2+\lambda-k+r}$. The
singularities arising in both cases are shown in
Table~\ref{table2}.
\begin{table}[t]
\caption{The singular contributions to $\widetilde{\sigma}(x)$
associated with the one-fold basic integrals. (Here $r$ is a
non-negative integer such that $5/2-k+r<0$.) }
{\begin{tabular}{l l l}
\\[-2mm]\hline\\[-2mm]
 $\qquad\qquad\qquad\qquad\qquad I^{1/2}_{k}$
 $\qquad\qquad\qquad\qquad$
          & $(x + a_j)_{+}^{5/2-k+r}$  $\qquad\qquad\qquad$ &
\\[3mm]
 $\qquad\qquad\qquad\qquad\qquad J^{1/2}_{k}$
 $\qquad\qquad\qquad\qquad$
          & $(-x - a_j)_{+}^{5/2-k+r}$ $\qquad\qquad\qquad$ &
\\[1mm]
\hline
\end{tabular}\label{table2}}
\end{table}

Thus, we have reduced integral (\ref{pap2}) to the sum of regular
and singular contributions. The regular contributions are
described by the absolutely convergent integrals and thereupon can
be numerically calculated. The singular contributions are
represented by the power-like distributions with regular weights.
So their further integration can be carried out with the aid of
formula (\ref{pap13}). It is worth noticing that the singularities
that remain at the level of hard-scattering cross-section
$\widetilde{\sigma}(x)$ are located on the physical thresholds
$x=0$ and $x = - a_i$.

So, we have discussed the piece of the hard-scattering
cross-section that initially possessed the double-resonant
structure. However, this piece includes not only the
double-resonant contributions but through the difference term
(\ref{pap7}) also the single-resonant and the non-resonant
contributions. Now we consider the piece of the cross-section
which initially possesses the single-resonant structure and which
is originated by the integral of the type (\ref{pap2}) without one
of the BW factors. The kinematic factor in this case is present as
before, but one of its variables, say, $x_j$ describes immediately
the invariant mass of the appropriate particles in the final
state. On the contrary, the $x_i$ variable describes first of all
the virtuality of the unstable particle and only then the
invariant mass of the appropriate stable particles. In this case
we do at first the MPT expansion of the BW factor which depends on
$x_i$ variable. Then the weight at each term of the resulting
expansion, we represent in the form of a finite power series in
$x_i$ with a remainder giving regular contribution. The latter
contribution may be reduced to an integral of the type
(\ref{pap15}), which is to be numerically calculated. The integral
$\d x_i$ of each term of the finite power series, we reduce to the
sum of the basic integrals (\ref{pap16}) and (\ref{pap17}). After
that the integrals $\d x_j$ we reduce to the sums of the integrals
of the type (\ref{pap18}). So we get again the sum of the regular
and singular contributions with the singular contributions
represented by the power-like distributions represented in
Table~\ref{table2}. Finally, the piece of the cross-section that
initially possesses the non-resonant structure, we consider in the
conventional fashion.

Now let us consider the interference contributions. In the case of
the interference between the contributions with the
single-resonant and the double-resonant structure, in the
cross-section the non-regular factors $(x_j \pm \i 0)^{-k}$ appear
instead of one of the BW factors ($k$ is a positive integer). This
factor is reduced via Sokhotsky formula to the sum of $PV
(x_j)^{-k}$ and $\delta_{k-1}(x_j)$. Thus we obtain contributions
that are similar to those which have been already discussed in the
case with the double-resonant structure. The other cases of the
interference are considered by similar fashion.

\section{The observable cross-section}\label{cross}
\setcounter{equation}{0}
\def\theequation{\ref{cross}.\arabic{equation}}

Having calculated the hard-scattering cross-section
$\widetilde{\sigma}$, we proceed to the calculation of the
convolution integral (\ref{not1}). In what follows we concentrate
on the singular contributions only. (Recall that the regular
contributions can be numerically calculated after the complete
definition of the weights.) In the dimensionless variables
integral (\ref{not1}) takes the form
\begin{equation}\label{cross1}
\sigma (s) = \frac{M^2}{2 M_1 M_2\,s}\,\int \d x' \> (4+x') \>
\theta(x'-x_{\mbox{\scriptsize min}}) \: \widetilde{\phi}(x',x) \>
\widetilde{\sigma}(x')\,.
\end{equation}
Here $\sqrt{s} = 2 M \,(1+ x/4)$, $\sqrt{s'} = 2 M \,(1+ x'/4)$,
$\sqrt{s_{\mbox{\scriptsize min}}} = 2 M \, (1+
x_{\mbox{\scriptsize min}}/4)$. The $\widetilde{\phi}(x',x)$ is
the flux function,
\begin{eqnarray}\label{cross2}
\widetilde{\phi}(x',x) \!&=&\! \beta_e
\left(4\!+\!x\right)^{2(1-\beta_e)}
\left(8\!+\!x\!+\!x'\right)^{\beta_e-1}
\left(x\!-\!x'\right)_{+}^{\beta_e-1} \nonumber\\[0.5\baselineskip]
 &-& \! \frac{1}{2} \>\beta_e  \left[ 1 +
 \left(\frac{4+x'}{4+x}\right)^{2} \right]\theta(x-x')\,.
\end{eqnarray}
In view of $\widetilde{\sigma}(x') \sim
\theta(x'\!+\!a_1\!+\!a_2)$ and $a_1 + a_2 < 4$, the integration
area in (\ref{cross1}) is restricted so that $x'>-4$. Consequently
only the factors $(x-x')_{+}^{\beta_e-1}$ and $\theta(x-x')$ are
unsmooth in the integration area.\footnote{In the strict sense,
the factor $\theta (x'-x_{\mbox{\tiny min}})$ in (\ref{cross1}) is
unsmooth, too. However if $x_{\mbox{\tiny min}}$ is the minimal
kinematically-allowed energy, this factor already is present in
$\widetilde{\sigma}(x')$ and therefore can be omitted in formula
(\ref{cross1}). In the case $x_{\mbox{\tiny min}} >
-(a_1\!+\!a_2)$, we suppose that $x_{\mbox {\tiny min}}$ does not
fall on the physical thresholds.} Owing to $\beta_e > 0$ both they
are integrable.

Further, we consider separately two cases, the first one when the
value of external variable $x$ is outside of the physical
thresholds $x=0$ and $x =-a_i$, and the second one when $x$ falls
precisely on one of the thresholds. In the former case the
singularities of $(x-x')_{+}^{\beta_e-1}$ and $\theta (x-x')$ do
not intersect with the singularities of $\widetilde{\sigma}(x')$.
So all singularities are isolated and integral (\ref{cross1}) can
be numerically calculated with the aid of formula (\ref{pap13}).
In the case when $x$ coincides with one of the thresholds, the
singularities intersect. However, owing to the relations $(\pm
x)_{+}^{\lambda_1} \: (\mp x)_{+}^{\lambda_2} = 0$ and $(\pm
x)_{+}^{\lambda_1} \: (\pm x)_{+}^{\lambda_2} = (\pm
x)_{+}^{\lambda_1 +\lambda_2}$ valid in the framework of the
analytical regularization, we get either zero contribution or a
single power-like distribution instead of the product of singular
distributions. So integral (\ref{cross1}) can again be numerically
calculated by means of formula (\ref{pap13}).

Unfortunately, in the vicinity of the thresholds the behavior of
the convolution integral is greatly unsmooth by the above
procedure, and therefore the above description cannot be
considered satisfactory. Namely the integral (\ref{cross1}) is
unboundedly increasing with $s$ approaching the thresholds
(although remaining finite precisely on the thresholds). For
definiteness let us consider the case $x \to 0$, which means $s
\to 4M^2$ in dimensional variables. In this case the dangerous
singularities in $\widetilde{\sigma}(x')$ are $(\pm
x')_{+}^\gamma$, where $\gamma$ is half-integer, because with $x
\to 0$ the position of singularity of $(x-x')_{+}^{\nu}$, where
$\nu$ is either $\beta_e\!-\!1$ or $0$, is approaching that of
$(\pm x')_{+}^\gamma$. At $\gamma + \nu < -1$ this implies the
increasing of the integral as $|x|^{1+\gamma+\nu}$. The similar
behavior takes place in the vicinities of the thresholds $x=-a_i$,
or $s=(\sqrt{s_{i\,\mbox{\scriptsize min}}}+M_j)^2$ in dimensional
variables.

In effect the mentioned behavior means a rapid divergence of the
asymptotic series when $s$ is approaching the thresholds, and thus
the expansion near thresholds becomes senseless. (Actually this
behavior have already been detected in \cite{N-tt} where also it
has been revealed that the area of bad behavior of the
cross-section is rather small.) Nevertheless, the situation may be
corrected by means of doing an additional expansion of the
cross-section $\sigma(s)$ near thresholds, namely the expansion in
powers of the distance between $s$ and the appropriate threshold.
For definiteness let us consider the physically important case
when $s$ is close to but a little bit lower of the threshold
$s=4M^2$. In the dimensionless variables this means
$x$~{\scriptsize $\lesssim$}~$0$. In this case in
$\widetilde{\sigma}(x')$ the singularity $(-x')_{+}^{\gamma}$ is
crucial. After changing the variable $x' \to x-\xi$, the
corresponding contribution to $\sigma (s)$ takes the form
\begin{equation}\label{cross3}
\sigma (s) \: \sim \: s^{-1}\int \d \xi \> \psi(\xi,x) \>
\xi_{+}^{\nu} \, (\xi\!-\!x)_{+}^{\gamma}.
\end{equation}
Here $\psi(\xi,x)$ is a regular contribution. Taking Taylor of
$(\xi\!-\!x)_{+}^{\gamma}$ in powers of $x$ (there is no necessity
to expand also the weight) gives the series
\begin{equation}\label{cross4}
\sum_{r=0,1,\,\cdots} \frac{(-x)^r}{r!} \;
\frac{\Gamma(\gamma+1)}{\Gamma(\gamma+1-r)} \;\, \xi_{+}^{\nu +
\gamma -r}
\end{equation}
instead of $\xi_{+}^{\nu} \, (\xi\!-\!x)_{+}^{\gamma}$. Since the
weight is regular in the vicinity of $\xi = 0$, the integral of
each term of this series can be numerically calculated. Truncating
the infinite series, we obtain an asymptotic estimate of $\sigma$
at small $x$. Notice that this procedure provides a continuity of
the result at deviating $s$ from threshold. The number of terms in
the truncated series should be chosen depending on the value of
the deviation of $s$ and of the current order of the expansion in
$\alpha$. Ultimately this number may be determined by the joining
of the solutions outside and inside the interval where the second
expansion of $\sigma(s)$ is made.

So, outside of the physical thresholds the asymptotic expansion of
the cross-section can be numerically calculated with the aid of
formula (\ref{pap13}). Near thresholds one should do the two-fold
expansion, one in powers of $\alpha$ and the other in powers of
the distance between $s$ and the appropriate threshold.

\section{Calculations up to NNLO}\label{NNLO}
\setcounter{equation}{0}
\def\theequation{\ref{NNLO}.\arabic{equation}}

In this section we discuss the characteristic features of the
application of the MPT within the NNLO approximation. For
methodological reasons we consider the expansion in the coupling
constant in a sequential manner. By the first we consider the
expansion of the BW factors, which by our definition let
constitute the MPT expansion in the narrow sense. From this point
it is convenient to trace more carefully the field of subsequent
calculations. Then we consider the expansion of the rest of
contributions to the cross-section.

First of all we consider contributions that possess the
double-resonant structure. In this case the LO in the narrow
sense, in accordance with (\ref{not10}), is determined by the
substitution of the $\delta$-functions (without the derivatives)
for the BW factors in formula (\ref{pap2}). Correspondingly, the
$\widetilde{\Phi}$ must be determined by the on-shell amplitude
multiplied by the branching factors, both calculated as a whole
with the two-loop precision. The soft massless-particles
contributions are to be taken into consideration by conventional
fashion. In particular, at calculating the amplitude the Coulomb
singularities are to be extracted in favor of the factor in the
cross-section. However, contrary to the wide belief, this factor
can be taken into consideration not in the form (\ref{not13})
only. In fact, it can be taken into consideration in the expanded
form, as well, though with certain modifications in subsequent
calculations. Up to the NNLO three terms of the expansion are
needed only, so we put
\begin{equation}\label{NNLO1}
1+\widetilde{\delta}_{c} = 1 + \frac{\kappa\,\alpha}{2\beta_0} +
\frac{1}{12}\left(\frac{\kappa\,\alpha}{\beta_0}\right)^2.
\end{equation}
The subsequent calculations can be carried out with replacing
$\beta_0$ by $\beta$ in the above formula, where $\beta \sim
\sqrt{x\!-\!x_1\!-\!x_2}$ is the off-shell velocity in the c.m.f.
Since the process in the end is considered on-shell, this
replacement should not affect the ultimate result. However at the
intermediate stage of calculations it implies an effective
modification of the kinematic factor. At the calculation of the
hard-scattering cross-section, the second and the third term of
(\ref{NNLO1}) manifest themselves through the basic integrals
$A^{\,0}_{1\,1}$ and $A^{-1/2}_{1\,1}$. From (\ref{calc17}) we
have $A^{\,0}_{1\,1}(x)=\theta(x)$ and $A^{-1/2}_{1\,1}(x)=
x_{+}^{-1/2}$. The convolution integral of such contributions
without problems can be calculated.

An alternative mode of taking into consideration Coulomb
singularities has been discussed in Sect.~\ref{not}. It implies a
regularization of the inverse powers of $\beta_0$ in (\ref{NNLO1})
by means of the special resummation. As applied to the second term
in (\ref{NNLO1}) this means a substitution
\begin{equation}\label{NNLO2}
\frac{\kappa\,\alpha}{2\beta_0} \;\to\; \mbox{r.h.s.~of
(\ref{not15})}\,.
\end{equation}
The substitution of this kind is inevitable at the calculation in
the NLO of the MPT in the narrow sense and in the higher orders,
because in this case not only the on-shell contributions of the
Coulomb factor must be taken into account but its derivatives,
too. Fortunately, in the NLO in the narrow sense, but remaining
within the NNLO in the general sense, it is sufficient to take
into consideration only the one-photon contributions to the
Coulomb factor, which is given by formula (\ref{not15}).
Simultaneously, the weight $\widetilde{\Phi}$ must be taken into
consideration off-shell and within the one-loop precision. The
soft massless-particles contributions that are not involved in the
Coulomb factor can be taken into consideration by means of
technique discussed in \cite{F2}.

In the NNLO in the narrow sense the weight $\widetilde{\Phi}$
should be taken off-shell in the Born approximation. The
contributions of soft massless particles are absent in this
approximation. Nevertheless, as long as the Coulomb factor makes
considerable contributions in the threshold region, its
contribution in the advance can be taken into account via formula
(\ref{not15}).

Further, we discuss the contributions that have single-resonant
structure. At the amplitude level they arise either due to the
process of inclusive single production of one of the unstable
particles, or due to non-factorizable corrections in the mode of
pair production. To the cross-section such processes can
contribute in the pure form without the interference, or in the
interference with the process of pair-production. Within the NNLO
in the general sense the mentioned contributions can appear in the
LO and NLO of the MPT in the narrow sense, and with the weight
calculated within the NLO of the conventional PT and in the Born
approximation, respectively. The unconventional counting of the
orders is explained by an occurrence of an additional power in
$\alpha$ when one of the unstable-particle propagators squared
disappears in the cross-section; cf.~(\ref{not5}) and
(\ref{not7}), and see discussion in \cite{N1,N2}. For the same
reason the interference between the non-resonant process and the
single- or double-resonant process contributes to the NNLO in the
general sense when the processes are calculated in the Born
approximation.

After the orders of the calculations are determined and the
corresponding weights are calculated, the further calculations are
to be made in accordance with the scheme of Sects.~\ref{pap} and
\ref{cross}. Here we notice only that in the NLO in the general
sense the following basic integrals are required: ${\tt
I}^{1/2}_{1}(x)$, $A^{1/2}_{1\,1}$, $B^{1/2}_{1\,1}$,
$B^{1/2}_{2\,1}$, $\overline{B\,}^{1/2}_{1\,1}$,
$\overline{B\,}^{1/2}_{1\,2}$. In the NNLO the additionally
required basic integrals are ${\tt J}^{1/2}_{1}(x)$, ${\tt
J}^{1/2}_{2}(x)$, $A^{1/2}_{2\,1}$, $A^{1/2}_{1\,2}$,
$A^{1/2}_{3\,1}$, $A^{1/2}_{1\,3}$, $C^{1/2}_{1\,1}$,
$C^{1/2}_{1\,2}$, $C^{1/2}_{2\,1}$, $C^{1/2}_{2\,2}$. The explicit
expressions for them follow immediately from the appropriate
formulas of Appendix~B. The distribution sense is strongly
required of the basic integrals $A^{1/2}_{3\,1}$,
$A^{1/2}_{1\,3}$, and $C^{1/2}_{2\,2}$.

\section{Discussion and conclusion}\label{disc}
\setcounter{equation}{0}
\def\theequation{\ref{disc}.\arabic{equation}}

We have constructed an asymptotic expansion in powers of the
coupling constant $\alpha$ of the cross-section for pair
production and decay of fundamental unstable particles. The
algorithm of the calculation of the coefficient functions of the
expansion is elaborated in an arbitrary order. The outcomes are
presented in the form of absolutely convergent integrals with
precisely definite structure. Certain components of the integrals
correspond to singular contributions in the amplitude and such
components are process-independent. The MPT is intended for
handling these singular contributions at the level of the
cross-section, while the regular contributions are to be
determined by means of conventional PT. The key moment of the MPT
is the interpretation of the singular contributions in the
distributions sense, so that they become integrable. In doing so,
the asymptotic property of the expansion is strictly maintained.
The exception is made for the Coulomb singularities, which are
considered extracted into a factor and represented in the
regularized form in the sense of conventional functions. However
the regularization implies no insertion of extrinsic parameter but
special resummation of intrinsic contributions
\cite{Fadin2}--\cite{Fadin4} So the asymptotic property of the
expansion in full measure is maintained.

In effect, the difference between the conventional PT and the MPT
appears at calculating the cross-sections and becomes apparent at
taking into consideration the resonant contributions of the
unstable particles. At comparing the MPT with the other approaches
intended for the description of the production and decay of
unstable particles, an essential moment distinguishing the MPT is
the complete expansion in powers of the coupling constant $\alpha$
of the resonant contributions of the unstable particles. In view
of this property one can expect the gauge cancellations at
calculating the cross-section in the framework of the MPT
approach.

The MPT method in the equal measure is applicable for the
calculation of the total cross-sections, the angular
distributions, and spin correlations. This follows immediately
from the fact that the angular and spin variables are not involved
in the BW factors. The calculation of the invariant-mass
distributions is workable in the MPT, too, but in conjunction with
some additional operations. Namely since the BW factors in the MPT
do not have the sense of conventional functions, an additional
integration of the BW factors with some weight is necessary. In
any case the integration should be made in view of the jets and
the contributions of unregistered photons in the final state.
Unfortunately, the convergence property of the MPT becomes worse
at the proceeding to the invariant-mass distributions. For this
reason we do not suggest the use of the MPT in this case, at least
in the considered above mode. For similar reason we do not suggest
the use of this mode of the MPT for the description of a generic
$2\to2$ process with a single resonance. Instead in these cases it
is better to use a modification of the MPT approach discussed in
\cite{N1,N2}, which is based on the secondary Dyson resummation in
the framework of MPT.

The question of the great importance in the case of pair
production of unstable particles is the near-threshold behavior of
the cross-section. Fortunately, this behavior can be calculated in
the MPT framework, as well, though through the double expansion of
the cross-section. Namely apart from the ordinary MPT expansion in
powers of $\alpha$, one should do expansion in powers of the
distance between $s$ and the appropriate threshold. The necessary
number of terms that must be retained in the latter expansion
depends on the value of the distance and on the current order of
the expansion in $\alpha$. Ultimately it is determined by the
claimed precision of the description.

The convergence properties of the MPT in the case of pair
production of unstable particles have been preliminaryly studied
in the framework of model calculations, in the above-threshold
region within the NLO \cite{N-tt}. (The modelling has concerned
the weights, but not the structure of the MPT itself.) The similar
calculations within the NNLO and with taking into consideration
the Coulomb singularities have been recently made, as well
\cite{prep}. In both cases a satisfactory behavior of outcomes has
been detected with essential improving of the description at the
proceeding to the NNLO. Numerical analysis in the threshold region
will be carried out elsewhere. The calculations of the completely
realistic processes will follow.

In conclusion, the basic result of the given work is that the very
existence is proved of an asymptotic expansion in powers of the
coupling constant of the cross-section for pair production and
decay of fundamental unstable particles. Furthermore, a scheme of
the calculation of the expansion in an arbitrary order has been
explicitly constructed and its specific ingredients have been
calculated. On the whole, the MPT method turns out to be a real
candidate for carrying out high-precision calculations necessary
for the realization of programs at colliders of next generation.

\section*{Acknowledgments}

The author is grateful to D.Bardin for the invitation to workshop
Calc2003, Dubna, where a basic part of this work was reported, and
to I.Ginzburg for promotional discussions.

\begin{flushleft}
\bf\large \underline{Appendix A}
\end{flushleft}
\setcounter{equation}{0}
\def\theequation{A\arabic{equation}}

Here we derive formula (\ref{pap4}) for the change of variables in
the $\delta$-function with derivatives. More precisely, we derive
a formula for the distribution $\delta_{n}\!\left(p(x)\right)$
resolved with respect to $x$ variable with
\begin{equation}\label{A1}
p(x) = x + b x^2\,,\qquad\qquad b \not= 0\,.
\end{equation}

At once we note that since the distribution $\delta_{n}(p)$ is
concentrated at $p=0$ (i.e.~gives zero contributions outside a
neighborhood of $p=0$), and since at the inverse mapping $p \to x$
the point $p=0$ turns into two points $x=0$ and $x =-1/b$, the
sought-for formula generally has the form
\begin{equation}\label{A2}
\delta_{n}(p) = \sum\limits_{k=0}^{n} \; z_k \; \delta_{k}(x) +
\sum\limits_{k=0}^{n} \; \zeta_k \; \delta_{k}(x+1/b) \,.
\end{equation}
Here $z_k$ and $\zeta_k$ are coefficients. However, our actual
interest is with the formula considered in a neighborhood of
$x=0$. For this reason we reduce the problem to the calculation of
the coefficients $z_k$ only.

First we note that in the case of small enough neighborhoods of
$x=0$ and $p=0$, formula (\ref{A1}) defines the one-to-one mapping
$p \leftrightarrow x$, with the inverse mapping determined by
formula
\begin{equation}\label{A3}
x = \frac{\sqrt{1+4bp}-1}{2b}\,.
\end{equation}
Further we note that by virtue of
\begin{equation}\label{A4}
\int \d x \; x^n\,\delta_{m}(x) = \delta_{nm}
\end{equation}
and in view of (\ref{A2}), the coefficients $z_k$ are determined
by the formula
\begin{equation}\label{A5}
z_k = \int \d x \; x^k\,\delta_{n}(p(x))\,,
\end{equation}
where $0 \le k \le n$ and the integration is carried out over the
above mentioned neighborhood of the point $x=0$. Making the change
of variable $x \to p$, we get
\begin{equation}\label{A6}
z_k = \int \d p \; \frac{1}{\sqrt{1+4bp}}\;
 \left( \frac{\sqrt{1+4bp}-1}{2b} \right)^k\,\delta_{n}(p)\,.
\end{equation}
An elementary calculation based on the formula for binomial
differentiation yields
\begin{equation}\label{A7}
z_k = \frac{(-4b)^n}{(2b)^k n!}\;\sum\limits_{r=0}^{k} \; {k
\choose r} \, (-)^{k+r} \,\frac{1}{\pi}\sin \!\left(\pi
\frac{r\!+\!1}{2} \right) \Gamma\!\left(\frac{r\!+\!1}{2} \right)
\Gamma\!\left(1+n-\frac{r\!+\!1}{2} \right).
\end{equation}
In this formula at odd $r$ the sine under the sum gives zero
contributions. So we can set $r=2m$, $m = 0, \dots, [k/2]$, where
$[k/2]$ is the integer part of $k/2$. After further manipulations
with the gamma-functions, we obtain
\begin{equation}\label{A8}
z_k = \frac{(-)^{k+n}\,b^n\,k!}{(2b)^k\,n!} \;
\sum\limits_{m=0}^{[k/2]} \; \frac{(-)^{m}}{m!}
\frac{(2n-2m)!}{(k-2m)!(n-m)!}\,.
\end{equation}
The calculation of the sum yields \cite{BMP}
\begin{equation}\label{A9}
z_k = {2n-k \choose n} \, (-b)^{n-k} \,.
\end{equation}

Now we substitute (\ref{A9}) into formula (\ref{A2}) considered
without the second sum. Making the change of variables $k \to
n-k$, we obtain
\begin{equation}\label{A10}
\delta_{n}(p(x)) = \sum\limits_{k=0}^{n} \; {n+k \choose n} \,
(-b)^{k}\; \delta_{n-k}(x)\,.
\end{equation}
Let us remember that the contributions located at $x = -1/b$ are
omitted in this formula.

A generalization to the case of presence of a common scale factor
in formula (\ref{A1}) is trivial and is practically implemented in
(\ref{pap4}).

\begin{flushleft}
\bf\large \underline{Appendix B}
\end{flushleft}
\setcounter{equation}{0}
\def\theequation{B\arabic{equation}}

In this Appendix we analytically calculate the single-subscript
basic integrals ${\tt I}^{\lambda}_{n}$, ${\tt J}^{\lambda}_{n}$
and the double-subscript basic integrals $A^{\,\lambda}_{n_1
n_2}$, $B^{\,\lambda}_{n_1 n_2}$, $C^{\,\lambda}_{n_1 n_2}$,
considered with positive parameters $a_i \,$ and non-integer
$\lambda$. In the single-subscript integrals that depend on one of
the parameters $a_i$, we omit the lower index ``$i$'' for
simplicity of the notation. In the case of the double-subscript
integrals we introduce $a=a_1+a_2$ and $n=n_1+n_2$.

\medskip

\begin{flushleft}{\it\bf Calculation of $\bf {\tt I}^{\lambda}_{n}$ }
\end{flushleft}

Let us discard the superfluous variable $x_j$ in formula
(\ref{pap16}) and omit the index ``$i$'' in the parameter $a_i$.
Then we come to an equivalent definition,
\begin{equation}\label{calc1}
{\tt I}^{\lambda}_{n}(x) = \int \d \xi \;\theta(\xi + a)\,
(x-\xi)_{+}^{\lambda}\,\delta_{n-1}(\xi)\,.
\end{equation}
Here $n$ is a positive integer, $a$ and $\lambda$ are continuous
parameters, $a > 0$ and $\lambda$ is non-integer. Recall that if
$\lambda$ is large enough, the $(x-\xi)_{+}^{\lambda}$ means
$\theta(x - \xi) \, (x-\xi)^{\lambda}$.

At first stage we consider $\lambda$ as large enough. By making
the change of variable $\xi \to x-\xi$ and resolving the
$\theta$-functions, we come to the equivalent formula for the
integral written in finite limits:
\begin{equation}\label{calc2}
 {\tt I}^{\lambda}_{n}(x) = \frac{\theta(x + a)}{(n-1)!}
 \int_{0}^{x+a} \d \xi \;\xi^{\lambda}\,\delta^{(n-1)}(\xi-x)\,.
\end{equation}
Notice that at $x < 0$ integral (\ref{calc2}) vanishes owing to
the strict positiveness of the argument in the $\delta$-function.
At $x>0$ the calculation by parts yields
\begin{equation}\label{calc3}
 {\tt I}^{\lambda}_{n}(x) = \frac{\theta(x + a)}{(n-1)!}\,
 (-)^{n-1}\,\theta(x)\left(x^{\lambda}\right)^{(n-1)} .
\end{equation}
Further we can omit factor $\theta(x+a)$, because at $a>0$ the
condition $x+a>0$ is automatically satisfied in the presence of
$\theta(x)$. After calculation of the derivatives, we obtain
\begin{equation}\label{calc4}
{\tt I}^{\lambda}_{n}(x) =
\frac{(-)^{n-1}\;\Gamma(1+\lambda)}{\Gamma(n)\,\Gamma(2+\lambda-n)}\;
\theta(x)\,x^{1+\lambda-n}\,.
\end{equation}

It is readily seen that at non-integer $\lambda$ this expression
is well defined for any $x$ except $x = 0$. At $x = 0$ the above
expression, generally, is ill defined. However eventually we do
not need a value of the function ${\tt I}^{\lambda}_{n}(x)$ at $x
= 0$, but a rule of its integration in a neighborhood of $x = 0$
with smooth weight. In the framework of analytical regularization
the $\theta(x)\,x^{1+\lambda-n}$ means the distribution
$x_{+}^{1+\lambda-n}$. So in the case of analytical regularization
we have
\begin{equation}\label{calc5}
{\tt I}^{\lambda}_{n}(x) =
\frac{(-)^{n-1}\;\Gamma(1+\lambda)}{\Gamma(n)\,\Gamma(2+\lambda-n)}\;
x_{+}^{1+\lambda-n}\,.
\end{equation}

With transiting to another regularization the result may change.
To make this more clear, let us consider an alternative derivation
of formula (\ref{calc5}) by basing on the direct calculation of
the integral in (\ref{calc1}). In this way at first we formally
obtain
\begin{equation}\label{calc6}
{\tt I}^{\lambda}_{n}(x) =
\frac{1}{(n\!-\!1)!}\,\frac{\d^{n-1}}{\d \xi^{n-1}}
\left[\theta(\xi + a)\theta(x - \xi) \,
 (x-\xi)^{\lambda}\right]_{\bigl| {\,\xi = 0} \bigr.}.
\end{equation}
Since with $a>0$ the differentiation of $\theta(\xi + a)$ gives
zero contribution at $\xi=0$, we omit this $\theta$-function. The
differentiation of other factors gives
\begin{equation}\label{calc7}
{\tt I}^{\lambda}_{n}(x)  = \frac{(-)^{n-1}}{(n\!-\!1)!} \;
\theta(x)\left(x^{\lambda}\right)^{(n-1)} + \;
\frac{(-)^{n-1}}{(n\!-\!1)!}\,
 \sum\limits_{r=1}^{n-1} {n\!-\!1 \choose r}\,\delta^{(r-1)}(x)
\left(x^{\lambda}\right)^{(n-r-1)}.
\end{equation}
In this formula the sum by definition equals zero if $n=1$. If
$n>1$, the contributions under the sum generally are ill-defined
and so they need a regularization. Generally, the different
regularizations lead to different outcomes. For instance, under
the cut-off regularization the sum takes the form of
$\sum_{r=1}^{n-1} C_r \,\delta^{(r)}(x)$, where $C_r$ are
non-vanishing coefficients. Under the analytical regularization
all coefficients $C_r$ vanish. Really, in the framework of the
analytical regularization formula (\ref{calc7}) must be considered
at first at $\lambda > n-2$, when all the terms in the r.h.s.~are
integrable. In this case all terms under the sum become zero in
view of the presence of $\delta^{(r-1)}(x)$. So only the first
term in (\ref{calc7}) survives. Having calculated the derivatives
in this term, we come to formula (\ref{calc5}) again.

\medskip

\begin{flushleft}{\it\bf Calculation of $\bf {\tt J}^{\lambda}_{n}$ }
\end{flushleft}

Proceeding on analogy to the previous case, we rewrite definition
(\ref{pap17}) in the form
\begin{equation}\label{calc8}
{\tt J}^{\lambda}_n(x) = \int \d \xi \;\theta(\xi + a)\,
(x-\xi)_{+}^{\lambda} \; PV \!\frac{1}{\xi^n}\,.
\end{equation}
Then, making the change of variable $\xi \to x-\xi$ and resolving
the $\theta$-functions, we get
\begin{equation}\label{calc9}
{\tt J}^{\lambda}_n(x) = \theta(x + a) \,(-)^n
 \int_{0}^{x+a} \d \xi \;
 \xi^{\lambda}\; PV \!\frac{1}{(\xi-x)^n}\,.
\end{equation}

Further let us consider an auxiliary integral with $b\!>\!0$,
\begin{equation}\label{calc10}
{\cal J}^{\lambda}_n(b,x) = \int_{\,0}^{\,b} \d \xi \;
\xi^{\lambda} \; PV \!\frac{1}{(\xi-x)^n}\,.
\end{equation}
At $\lambda>n-1$ and $x\not=0$, $x\not=b$ this integral can be
calculated by means of reducing the improper integral to the
derivatives of the absolutely convergent integral:
\begin{equation}\label{calc11}
\int_{\,0}^{\,b} \!\!\! \d \xi \; \xi^{\lambda} \; PV \!
\frac{1}{(\xi-x)^n} = - \,\frac{1}{(n\!-\!1)!} \; \frac{\d^n}{\d
x^n} \!\int_{\,0}^{\,b} \!\!\! \d \xi \; \xi^{\lambda} \;
\ln|\xi-x| .
\end{equation}
The integral in the r.h.s.~can be explicitly calculated. At $b>x$
and non-integer $\lambda>-1$, we have
\begin{eqnarray}\label{calc12}
&\displaystyle \int_{\,0}^{\,b} \d \xi \; \xi^{\lambda} \;
\ln|\xi-x|
  \;=\;  \frac{b^{1+\lambda}}{1\!+\!\lambda} \;
\left[ \ln(b-x) - \frac{1}{1+\lambda}\,
 F\!\!\left(1,-\lambda-1;-\lambda; \frac{x}{b}\right)\right]\quad&
 \\[0.7\baselineskip]
&\displaystyle +\;
\frac{\Gamma(1\!+\!\lambda)\Gamma(\!-\lambda)}{1\!+\!\lambda}\!
 \left[-\!\cos(\pi\lambda) \, \theta(x)\,x^{1+\lambda}\!+\!
 \theta(\!-x)\,(\!-x)^{1+\lambda}\right] .&\nonumber
\end{eqnarray}
Hereinafter $F$ means the hypergeometric function $_{2}F_{1}$
\cite{Bateman}. Substituting (\ref{calc12}) into (\ref{calc11})
and carrying out the differentiation, we get
\begin{eqnarray}\label{calc13}
&\displaystyle {\cal J}^{\lambda}_n(b,x) \;=\;
\frac{b^{1+\lambda-n}}{1+\lambda-n} \;
F\!\!\left(n,n-\lambda-1;n-\lambda; \frac{x}{b}\right)&
\\[0.7\baselineskip]
&\displaystyle +\;
\frac{\Gamma(1+\lambda)\,\Gamma(n-\lambda-1)}{\Gamma(n)}
\left[(-)^{n-1}\cos(\pi\lambda)\,\theta(x)\,x^{1+\lambda - n}
 +\theta(-x)\,(-x)^{1+\lambda - n}\right] .\nonumber
\end{eqnarray}

Comparing (\ref{calc9}) with (\ref{calc10}) and (\ref{calc13}), we
come to the formula:
\begin{eqnarray}\label{calc14}
&\displaystyle {\tt J}^{\lambda}_n(x) =
 (-)^{n} \, \frac{(x + a)_{+}^{1+\lambda-n}}{1+\lambda-n} \,
F\!\!\left(n,n\!-\!\lambda\!-\!1;n\!-\!\lambda; \frac{x}{x +
a}\right)&
\\[0.7\baselineskip]
&\displaystyle +\; \theta(x+a) \,
\frac{\Gamma(1+\lambda)\,\Gamma(n-\lambda-1)}{\Gamma(n)}
\left[-\cos(\pi\lambda)\,x_{+}^{1+\lambda - n}
 + (-)^n\,(-x)_{+}^{1+\lambda - n}\right] .\nonumber
\end{eqnarray}
We see that ${\tt J}^{\lambda}_n(x)$ has, in general, two singular
points, the $x=0$ and $x=-a$. At $x=0$ the singularity is
controlled by the second term in (\ref{calc14}). To find the
behavior at $x \to -a$, we take advantage of the
analytic-continuation formula for the hypergeometric function with
inverse argument \cite{Bateman}. In this way at $x<0$ we obtain
\begin{eqnarray}\label{calc15}
{\tt J}^{\lambda}_n(x) = \frac{(x+a)_{+}^{1+\lambda}}{(-x)^n
\,(1+\lambda)} \,F\!\left(\!n,1+\lambda;2+\lambda; \frac{x +
a}{x}\right).\nonumber\\&&
\end{eqnarray}
This formula is equivalent  to (\ref{calc14}) at $-a<x<0$. With
$1+\lambda>0$ the above expression is regular at $x + a \to 0$.

\medskip

\begin{flushleft}{\it\bf Calculation of $A^{\,\lambda}_{n_1 n_2}$ }
\end{flushleft}

Let us consider integral (\ref{pap10}) as the iterated one, and
let us at first calculate integral in $x_2$. With the aid of
(\ref{calc1}) and (\ref{calc5}) we obtain
\begin{equation}\label{calc16}
A^{\,\lambda}_{n_1 n_2} (x) =
 \frac{ (-)^{n_2\!-\!1} \; \Gamma(1+\lambda) }
      { \Gamma(n_2) \, \Gamma(2\!+\!\lambda\!-\!n_2) }\,
      \int\!\d x_1 \; \theta(x_{1}\!+\!a_{1})
 \left( x-x_1 \right)^{1+\lambda-n_2}_{+} \,
 \delta_{n_1\!-\!1}(x_1)\,.
\end{equation}
The similar calculation of the remaining integral yields
\begin{equation}\label{calc17}
A^{\,\lambda}_{n_1 n_2} (x) = \frac{(-)^n\;
\Gamma(1+\lambda)}{\Gamma(n_1)\,\Gamma(n_2)\,
\Gamma(3\!+\!\lambda\!-\!n)} \; x^{2+\lambda-n}_{+}\,.
\end{equation}
Here $n=n_1+n_2$. The result is symmetric and does not depend on
the (positive) para\-meters $a_i$.

\medskip

\begin{flushleft}{\it\bf Calculation of $\bf B^{\,\lambda}_{n_1 n_2}$}
\end{flushleft}

As in the previous case, we consider integral (\ref{pap11}) as the
iterated one and at first calculate integral $\d x_2$. With the
aid of (\ref{calc1}) and (\ref{calc5}) we get
\begin{equation}\label{calc18}
B^{\,\lambda}_{n_1 n_2} (x) = \frac{ (-)^{n_2\!-\!1} \;
\Gamma(1+\lambda)}
      { \Gamma(n_2) \, \Gamma(2\!+\!\lambda\!-\!n_2) }\,
 \int\!\d x_1 \;\theta(x_{1}\!+\!a_{1})
 \left( x-x_1 \right)^{1+\lambda-n_2}_{+}\,
 PV \!\frac{1}{x_1^{n_1}}\,.
\end{equation}
Then, with the aid of (\ref{calc8}) and (\ref {calc14}) we obtain
\begin{eqnarray}\label{calc19}
&\displaystyle B^{\,\lambda}_{n_1 n_2}(x) \;=\; (-)^{n-1}
 \frac{\Gamma(1+\lambda)}{\Gamma(n_2)\,\Gamma(2+\lambda-n_2)}\,
 \frac{(x+a_1)_{+}^{2+\lambda-n}}{2+\lambda-n}&
  \\[0.5\baselineskip]
&\displaystyle \times \;
F\!\left(\!n_1,n\!-\!\lambda\!-\!2;n\!-\!\lambda\!-\!1;\frac{x}{x+a_1}
\right)&
 \nonumber\\[0.5\baselineskip]
&\displaystyle -\;\; \theta(x+a_1)\,
 \frac{\Gamma(1+\lambda)\,\Gamma(n-\lambda-2)}
 {\Gamma(n_1)\Gamma(n_2)}
 \left[\cos(\pi \lambda)\,x_{+}^{2+\lambda-n}
 +(-)^{n}\,(-x)_{+}^{2+\lambda-n}\right] .&\nonumber
\end{eqnarray}
The result is independent from the order of the calculation of
integrals, which may be verified by direct calculations. Notice
also that (\ref{calc19}) is independent from the parameter $a_2$.
The conjugate integral $\overline{B\,}$$^{\,\lambda}_{n_1 n_2}$ is
derived by the substitution $\{n_1,n_2,a_1\} \to \{n_2,n_1,a_2\}$.

In general, there are two singular points in $B^{\,\lambda}_{n_1
n_2}(x)$, the $x=0$ and $x+a_1=0$. At $x \to 0 $ the singular
behavior is completely controlled by the second term in
(\ref{calc19}). The behavior at $x+a_1 \to 0$ can be found on the
basis of the representation of the hypergeometric function $F$ in
the form with inverse argument \cite{Bateman}. At $x<0$ we obtain:
\begin{equation}\label{calc20}
B^{\,\lambda}_{n_1 n_2}(x) = (-)^{n-1}
\frac{\Gamma(1+\lambda)\:(x+a_1)_{+}^{2+\lambda-n_2}}
{\Gamma(n_2)\,\Gamma(3+\lambda-n_2)\:(-x)^{n_1}}\,
F\!\left(\!n_1,2\!+\!\lambda\!-\!n_2;3\!+\!\lambda\!-\!n_2;\frac{x+a_1}
{x} \right).
\end{equation}
At $2+\lambda-n_2>0$ this expression is regular as $x \to -a_1$.
At $2+\lambda-n_2 < 0 $ it contains singularity
$(x+a_1)_{+}^{2+\lambda-n_2}$ and also the associated set of
singularities $(x+a_1)_{+}^{2+\lambda-n_2+r}$, where $r$ is a
positive integer such that $2+\lambda-n_2+r < 0$.

\medskip

\begin{flushleft}{\it\bf Calculation of $\bf C^{\,\lambda}_{n_1 n_2}$ }
\end{flushleft}

The calculation of $C^{\,\lambda}_{n_1 n_2}$ is a more complicated
task. We begin with the observation that at large enough $\lambda$
the singularities in the integrand in (\ref{pap12}) do not
intersect. So the integral is well defined as the iterated one and
theoretically can be calculated on the basis of the above results.
However in view of complexity of the intermediate expressions, the
analytical calculation in this way unlikely is efficient. An
efficient method from the very beginning should allow for
symmetrical properties of the integral, which are evident when
considering the integral as the double one. (Notice that a double
integral of the $PV$-poles and of the $\delta$-functions may be
defined through the representation of these singular functionals
as improper limits of the regular functionals, see for instance
\cite{Gelfand} and \cite{Mik}.)

So let us consider integral (\ref{pap12}) as the double one. For
the reasons that become clear later, we begin calculations with
adding and subtracting a certain term proportional to $A_{n_1
n_2}^{\,\lambda}$. Namely we rewrite (\ref{pap12}) in the form:
\begin{eqnarray}\label{calc21}
C_{n_1 n_2}^{\,\lambda} (x) &=&
 (-)^{n+1} \pi^2 A_{n_1 n_2}^{\,\lambda}
 + \int\!\!\!\int\!\d x_1 \, \d x_2 \;
 \theta(x_{1} + a_{1})\,\theta(x_{2} + a_{2})\,
 (x\!-\!x_1\!-\!x_2)_{+}^{\lambda}
 \nonumber\\[0.5\baselineskip]
 &\times&
 \left\{PV \!\frac{1}{x_1^{n_1}}\;PV \!\frac{1}{x_2^{n_2}}
 \;+\; (-)^{n} \pi^2 \,
 \delta_{n_1-1}(x_1) \, \delta_{n_2-1}(x_2)
 \right\}. \qquad
\end{eqnarray}
Note that the integration in (\ref{calc21}) goes over a simplex.
So for providing symmetric integration, we make a shift $x_i \to
x_i - a_i$ and then a transition to the cone variables $x_1 + x_2
= \xi$, $x_1 - x_2 = 2 \eta$. After that, formally representing
the double integral again as the iterated one, we obtain
\begin{eqnarray}\label{calc22}
&\displaystyle C_{n_1 n_2}^{\,\lambda} (x) \;=\;
 (-)^{n+1} \pi^2 A_{n_1 n_2}^{\,\lambda} + \;
 \theta(x+a) \, (-)^{n_2}\!\!\! \int\limits_0^{x+a} \!
 \d \xi \; (x + a - \xi)^{\lambda}  &
 \\
& \displaystyle \times \int\limits_{-\xi/2}^{\xi/2} \d \eta
 \left\{
 PV \frac{1}{(\eta + \xi/2 - a_1)^{n_1}}\,\,
 PV \frac{1}{(\eta - \xi/2 + a_2)^{n_2}}
 \right.& \nonumber\\
& \displaystyle
 \qquad\qquad\qquad\qquad + \; \Biggl. (-)^{n+1} \pi^2 \;
 \delta_{n_1-1}(\eta + \xi/2 - a_1)\;
 \delta_{n_2-1}(\eta - \xi/2 + a_2)
 \biggr\} .&\nonumber
\end{eqnarray}
Unfortunately, in this formula the expression in the curly
brackets, generally, is ill-defined. Really, with $x>0$ the
variable $\xi$ may become equal to $a=a_1\!+\!a_2$ with the
consequence that there appear the intersecting singularities.
However this behavior, actually, is an artefact of the transition
to cone variables and the problem may be removed on the basis of
the following relation \cite{math} (which, in turn, is a
generalization of the relation obtained in a simpler case
\cite{Musk}):
\begin{eqnarray}\label{calc23}
& \displaystyle PV \!\frac{1}{(z-z_1)^{n_1}} \; PV
\!\frac{1}{(z-z_2)^{n_2}} + (-)^{n+1} \pi^2 \;
 \delta_{n_1-1}(z-z_1)\delta_{n_2-1}(z-z_2) &
\\[0.3\baselineskip]
& \displaystyle = \;
 \sum\limits_{r=0}^{n_1-1} \mbox{${n_2+r-1 \choose r}$}\,(-)^r\,
 PV \!\frac{1}{(z_1-z_2)^{n_2+r}}\>
 PV \!\frac{1}{(z-z_1)^{n_1-r}}
 &\nonumber \\
& \displaystyle + \;
 \sum\limits_{r=0}^{n_2-1} \mbox{${n_1+r-1 \choose r}$}\,(-)^r\,
 PV \!\frac{1}{(z_2-z_1)^{n_1+r}}\>
 PV \!\frac{1}{(z-z_2)^{n_2-r}}\,.
 &\nonumber
\end{eqnarray}
The meaning of this relation is as follows. Let us assume that
some integral with a weight of the expression in the l.h.s.~is to
be calculated at first in $z$ and then in $z_1$ or $z_2$. Then the
calculation of the integral in this sequence should be carried out
with the substitution of the expression in the r.h.s~in this
formula. In our case, we need a projection of the relation
(\ref{calc23}) obtained by substitutions $z=\eta$, $z_1 =
-\xi/2+a_1$, $z_2 = \xi/2-a_2$. After the substitutions, the
l.h.s.~in (\ref{calc23}) becomes the same expression as the in the
curly brackets in (\ref{calc22}). So the initial integral can be
converted to the form of (\ref{calc22}) with substituting the
following expression for the expression in the curly brackets:
\begin{eqnarray}\label{calc24}
& \displaystyle \sum\limits_{r=0}^{n_1-1} \!\!
  \mbox{$ {n_2+r-1 \choose r}$} (-)^r
  PV \! \frac{1}{(a\!-\!\xi)^{n_2+r}}\>
  PV \! \frac{1}{(\eta\!+\!\xi/2\!-\!a_1)^{n_1-r}}&  \\
& \displaystyle + \;
  \sum\limits_{r=0}^{n_2-1} \!\!
  \mbox{$ {n_1+r-1 \choose r}$} (-)^r
  PV \! \frac{1}{(\xi\!-\!a)^{n_1+r}}\>
  PV \! \frac{1}{(\eta\!-\!\xi/2\!+\!a_2)^{n_2-r}}\,.&\nonumber
\end{eqnarray}
A consideration of completely solvable example \cite{math}
independently shows that the substitution of (\ref{calc24}) solves
the problem of the ghost singularities emerging at the transition
to the cone variables.

So in the integral $\d \eta$ in (\ref{calc22}) we gain the
contribution of a sum of isolated $PV$-poles instead of
intersecting singularities. The integral of the poles, we
calculate with the aid of the formulas
\begin{equation}\label{calc25}
\int_{a}^{b} \d \eta \; PV \frac{1}{\eta-\xi} \; = \; \ln|b-\xi| -
\ln|a-\xi|\,, \qquad
\end{equation}
\begin{equation}\label{calc26}
\int\limits_{a}^{b} \d \eta \; PV \!\frac{1}{(\eta-\xi)^n}  =
-\,\frac{1}{n\!-\!1} \left[PV\frac{1}{(b-\xi)^{n\!-\!1}} - PV
\!\frac{1}{(a-\xi)^{n\!-\!1}}\right].
\end{equation}
After the calculation of the integral $\d \eta$, we make the
change of variables $\xi \to x+a-\xi$ in the remaining integral
$\d \xi$ and simultaneously the change $r \to n_i - r -1$ in
(\ref{calc24}). This yields ($a = a_1+a_2$)
\begin{eqnarray}\label{calc27}
&& C_{n_1 n_2}^{\,\lambda} (x) \;=\;
 (-)^{n+1} \pi^2 A_{n_1 n_2}^{\,\lambda}(x) \\[0.6\baselineskip]
&+&\!\! \theta(x\!+\!a)
 \frac{(-)^{n-1}\Gamma(n\!-\!1)}{\Gamma(n_1)\,\Gamma(n_2)}
 \int\limits_{0}^{x+a} \!\d \xi \; \xi^{\lambda}\!
 \left(\ln\!\left|\frac{\xi\!-\!x\!-\!a_1}{a_1}\right|
 + \ln\!\left|\frac{\xi\!-\!x\!-\!a_2}{a_2}\right|\,
 \right)\! PV \! \frac{1}{(\xi\!-\!x)^{n-1}}
\nonumber\\[0.3\baselineskip]
&+&\!\! \theta(x+a) (-)^{n-1}\!
 \left[
  \sum\limits_{r=1}^{n_1-1} \mbox{$ {n-r-2 \choose n_2-1}$}
     \,\frac{a_1^{-r}}{r}
 + \!\! \sum\limits_{r=1}^{n_2-1} \mbox{$ {n-r-2 \choose n_1-1}$}
     \,\frac{a_2^{-r}}{r}
 \right]\!
 \int\limits_{0}^{x+a} \!\!\d \xi \; \xi^{\lambda}\;
  PV \!\frac{1}{(\xi\!-\!x)^{n-r-1}} \nonumber\\
&+&\!\! \theta(x+a) (-)^{n}
 \sum\limits_{r=1}^{n_1-1} \mbox{$ {n-r-2 \choose n_2-1}$}
 \,\frac{1}{r}
 \int\limits_{0}^{x+a} \d \xi \; \xi^{\lambda}\;
  PV \!\frac{1}{(\xi-x)^{n-r-1}}\;PV \!\frac{1}{(\xi-x-a_2)^{r}}
\nonumber
%\\
\end{eqnarray}
\begin{eqnarray}
&+&\!\! \theta(x+a) (-)^{n}
 \sum\limits_{r=1}^{n_2-1} \mbox{$ {n-r-2 \choose n_1-1}$}
 \,\frac{1}{r}
  \int\limits_{0}^{x+a} \d \xi \; \xi^{\lambda}\;
  PV \!\frac{1}{(\xi-x)^{n-r-1}}\;PV
\!\frac{1}{(\xi-x-a_1)^{r}}\,.\nonumber
\end{eqnarray}
In this formula if the upper bound in any sum is less than the
lower bound, then by definition the sum is zero.

Let us discuss the obtained outcome. First we note that, actually,
we have calculated the integral in the third term in
(\ref{calc27}) while calculating the basic integral ${\tt
J}^{\lambda}_{n}$. The integrals in the two last terms can be
easily calculated, as the singularities of the $PV$-poles do not
coincide at $a_i \not= 0$. So the only problem integral remains in
the second term, which contains logarithm contributions. To
calculate it, we subtract and add to each logarithm the $n-1$ of
the first terms of its expansion in powers of $\xi-x$, so that to
represent the logarithm in the form
\begin{equation}\label{calc28}
\ln\left|\frac{\xi-x-a_i}{a_i}\right| =
 \left\{\,
  \ln\left|\frac{\xi-x-a_i}{a_i}\right| \;+\,
  \sum\limits_{r=1}^{n-2} \frac{1}{r}
  \left(\frac{\xi-x}{a_i}\right)^r \,
 \right\} - \sum\limits_{r=1}^{n-2} \frac{1}{r}
  \left(\frac{\xi-x}{a_i}\right)^r .
\end{equation}
Here the contribution in the curly brackets is of order
$O((\xi-x)^{n-1})$ as $\xi-x \to 0$. Therefore the appropriate
contribution to the integral is regular in the vicinity of $\xi =
x$. The contribution of the last term in (\ref{calc28}) is reduced
to the sum of integrals considered at the calculation of ${\tt
J}^{\lambda}_{n}$.

After carrying out all above mentioned calculations, we come to
the following result:
\begin{eqnarray}\label{calc29}
C_{n_1 n_2}^{\,\lambda} (x) \;&=&\; (-)^{n+1} \pi^2 A_{n_1
n_2}^{\,\lambda}(x)\> +
\;\frac{(-)^{n-1}\,\Gamma(n\!-\!1)}{\Gamma(n_1)\,\Gamma(n_2)}
   \int \d \xi \;\theta(x+a-\xi)\,\xi_{+}^{\lambda}\;
   \chi_n(\xi,x)\,, \nonumber\\
\end{eqnarray}
\begin{equation}\label{calc30}
\chi_n(\xi,x) = \frac{1}{(\xi-x)^{n-1}}
 \left\{\,\ln\left|\frac{\xi-x-a_1}{a_1}\right|
      \,+ \sum\limits_{r=1}^{n-2} \,\frac{1}{r}
        \left(\frac{\xi-x}{a_1}\right)^{r}\! + \Bigl(a_1 \to a_2
        \Bigr)
 \right\}.
\end{equation}
Since $\chi_n(\xi,x)$ by the construction is a regular function,
integral (\ref{calc29}) is absolutely convergent at $\lambda >
-1$. In this case the singularity of $C_{n_1 n_2}^{\,\lambda}(x)$
is contained only in the $A_{n_1 n_2}^{\,\lambda}(x)$ which is
present in (\ref{calc29}).

\end{document}